\documentclass[preprint,aps,tightenlines,floatfix,showpacs]{revtex4}

\usepackage{graphics}
\usepackage{bm}
\usepackage{amsmath}
\usepackage{amssymb}
\begin{document}

\title{Complementary reaction analyses and the isospin mixing of the
4$^-$ states in ${}^{16}$O.}

\author{K. Amos} 
\email{amos@physics.unimelb.edu.au}
\author{S. Karataglidis}
\email{kara@physics.unimelb.edu.au}
\author{Y. J. Kim}
\email{yjkim@cheju.ac.kr}
\affiliation{School of
Physics, The University of Melbourne, Victoria 3010, Australia}

\date{\today}

\begin{abstract}
Data from the inelastic scattering of electrons, and of intermediate
energy  protons and   pions  leading  to  ``stretched'' configuration
4$^-$ states near  19  MeV excitation in ${}^{16}$O as well as from 
charge exchange $(p,n)$ scattering to an  isobaric analogue (4$^-$) 
state in  ${}^{16}$F  have been analyzed to ascertain the degree of 
isospin mixing contained within  those states and of the amount  of
$d_{5/2}-p_{3/2}^{-1}$  particle-hole  excitation  strength  they 
exhaust. The electron and proton scattering data have been analyzed 
using microscopic models of the structure and reactions, with details
constrained by analyses of elastic scattering data. 
\end{abstract} 

\maketitle

%%%%%%%%%%%%%%%%%%%%%%%%% INTRODUCTION %%%%%%%%%%%%%%%%%%%%%%%%
\section{Introduction}

           The nucleus is an unique environment in that the strong, 
electromagnetic, and weak interactions all contribute in ways  that 
are manifest in its static and dynamic attributes. Furthermore,  as 
nuclear systems  display a rich  array  of properties,  it  is  not 
surprising that diverse reaction studies, using a variety of probes
and processes, are required to provide  a  range  of  complementary 
information before a detailed understanding of nuclear structure is 
possible.         A requirement is that the reaction mechanisms  of 
importance in each use (types and specifications) are well known.

With   the  electromagnetic  interaction,  for  both $\gamma$-decay 
probabilities and electron scattering form factors that supposition
is well founded~\cite{Ub71},     at least for momentum transfers to 
$\sim$ 3 fm$^{-1}$.   At higher  momentum  transfer  values,  meson
exchange current   (MEC)   corrections  may have noticeable effect 
particularly on transverse form factors. For electric  transitions,  
other  studies~\cite{Fr84,Fr85}  have shown how effective operators 
can be defined to  account  for MEC effects. With large basis shell 
model calculations of structure those corrections have  led to good 
agreement with $E2$ transition data~\cite{Ka95,Am00}.        Magnetic 
transverse form factors also are affected  by  MEC,  but  as  their 
evaluations explicitly involve the nuclear currents, individual MEC 
diagrams must be evaluated.      While the one pion exchange current
(seagull and pion in flight)  contributions  have  little effect at 
small linear momentum transfer values in scattering to states 
involving large angular momentum change,   they 
have some effect at higher momentum transfer  values  and  do so in 
the case of the $M4$ values in ${}^{16}$O~\cite{Hy87},   though   the
effects are still minor and lie within  vagaries  of  the choice of
single nucleon bound state wave functions~\cite{Cl88}.

         Data from the inelastic pion scattering from nuclei are of 
particular interest, especially for pions with energies that ensure
the $\Delta$ resonance dominance of the underlying pion-nucleon ($\pi N$)
interactions effecting the transitions.  The interest stems largely
from the distinctive relative transition strengths of  $\pi^+$  and 
$\pi^-$ interactions with protons and neutrons.   As a consequence, 
$\pi^{(\pm)}$    inelastic  scattering  cross  sections  should  be 
sensitive to the  isospin of (and so isospin mixing in) the nuclear 
states.  But the specifics of the $\pi N$ $g$-matrices in nuclear 
matter are not well known.   

   In contrast, credible nucleon-nucleon ($NN$) $g$-matrices in the 
nuclear medium have been formed for an energy of an incoming nucleon
from  25 to 250 MeV. Using effective interactions that map to those
$g$-matrices,  microscopic model (nonlocal) nucleon-nucleus optical 
potentials have been determined with which very good predictions of 
cross  sections and spin observables from the elastic scattering of 
nucleons with energies in that range and  from nuclei that span the
mass table~\cite{Am00}.    Such model evaluations of proton elastic 
scattering   from   ${}^{6,8}$He~\cite{La01},  and  from 
${}^{208}$Pb~\cite{Ka02,Ka05}, gave estimates for the neutron  skin 
thickness of each nuclei.    Specifically,  calculations of nucleon 
elastic scattering were made  using  complex,  spin-dependent,  and
nonlocal  optical  potentials  formed  by  folding   an   effective 
two-nucleon ($NN)$ interaction with  nuclear state density matrices 
from a large basis space shell model calculation.    That effective  
$NN$ interaction accurately maps to the $NN$  $g$-matrices half off 
the energy shell, solutions of the Brueckner-Bethe-Goldstone  (BBG) 
equations for nuclear matter found for diverse Fermi momenta to 1.6 
fm$^{-1}$ and built upon the BonnB $NN$ potentials.    The nonlocal 
optical  potentials  so  formed  were  used  also  to  specify  the 
continuum  waves  in  a Distorted Wave Approximation (DWA) study of 
inelastic scattering.  The effective $NN$ interaction was also used 
as the transition operator in those DWA calculations as was a large
basis  space  shell  model  spectroscopy  for  the  nuclear states.    
Remarkably  good  comparisons with  data  resulted,   requiring  no 
arbitrary renormalization  or adjustments   to  bound  state  wave 
functions~\cite{Am00}.   The inelastic scattering results so found,
cross  correlated  extremely well with the electron scattering form 
factor predictions of the selected model of nuclear structure.

      The study of transitions to the 4$^-$ states in ${}^{16}$O is 
particularly interesting.   First, the dominant spectroscopy of the 
residual    states    should    be     particle-hole    excitations 
($d_{5/2}-p_{3/2}^{-1}$) upon the ${}^{16}$O ground state.   If an
essentially closed shell model for that ground state is considered, 
only two such states should exist; one each with isospin ($T$) of 0 
and 1.   But there are four known  4$^-$  states  in  the   adopted 
spectrum~\cite{Pe93}  in  the  vicinity  of  19 MeV excitation; the 
three lowest having observable excitation strength from the  ground 
with one or more of the scattering of electrons, pions and protons.
Two of those are dominantly isoscalar.   The third is classified as 
an isovector state whose analogue in ${}^{16}$F  has  been  seen in 
charge    exchange    $(p,n)$    experiments~\cite{An79,Ma82,Oh82}. 
Furthermore, electron scattering (transverse) form factors for  two
of  those  states have been measured and analyzed~\cite{Hy87,Cl88}, 
$\pi^{(\pm)}$ inelastic  scattering cross sections  for energies in 
the $\Delta$ resonance region have  been  studied~\cite{Ho80},  and  
the cross section from inelastic scattering and cross section and
analyzing power from charge exchange scattering  of 
135 MeV protons have  been reported~\cite{He79,Ma82}.

       In the next section, the spectroscopy of the $4^-$ states is 
discussed giving a particle-hole model for their excitation   (from
the ground) and also a  prescription  to  study  configuration  and
isospin mixing among them.        Then in Sec.~\ref{Theory} a brief 
outline of details of  the  calculations  of  electron,  pion,  and 
proton inelastic scattering and  of charge exchange  scattering  to 
the isobaric analogue state (IAS) state in ${}^{16}$F is presented.     
The results of those 
calculations are then shown in  Sec.~\ref{Results}  and conclusions 
we draw follow in Sec.~\ref{Conclu}.

%%%%%%%%%%%%%%%%%  SPECTROSCOPY %%%%%%%%%%%%%%%%%%%%%%%%%%%%%%
\section{ The spectroscopy of the 4$^-$ states}
\label{spectroscopy}

  Simple shell model calculations of ${}^{16}$O readily predict two 
4$^-$ states in the spectrum. These are the isoscalar and isovector 
combinations of the ``stretched'' particle-hole excitations upon the
ground  state  with  a  $p_{3/2}$  proton(neutron)  moving  to  the 
$d_{5/2}$ orbit.         The best of such simple model calculations
~\cite{Mi75} predict excitation energies for those states of 19 MeV 
in accord with observation.  The transition strength for excitation 
of such ``stretched'' states however,   depends upon the fractional
occupancies of the $p_{3/2}$ and $d_{5/2}$  orbits  in  the  ground 
state.  So one must have a good description for the ground state as 
well. The fractional occupancies   can  have  a marked  effect upon 
the  normalization  of  cross-section peak values as was noted in a 
study of the excitation of  the isoscalar/isovector  pair  of 6$^-$ 
states in ${}^{28}$Si~\cite{Sm78}  wherein it  was shown that using 
a  large  basis  projected Hartree-Fock specification of the ground 
state could halve the strength of transition obtained from using  a 
packed shell model description.

        Diminution of particle-hole excitation strength may be 
anticipated with any ``stretched'' states isospin pair. But in the case
of ${}^{16}$O, there are three strongly excited 4$^-$ states in the
region of 19 MeV excitation.      Specifically they have excitation
energies of  17.775,  18.977 and  19.808 MeV and (dominant) isospin
values of 0, 1, and 0 respectively~\cite{Pe93}.   There is a fourth 
listed now~\cite{Pe93} at an excitation of  20.5~MeV to which we do
not have scattering data.      Additional isoscalar negative parity
states are not unexpected though as such are noted in the  evaluated 
spectrum as well as being predicted by shell model calculations that
allow $3p-3h$ excitations~\cite{Mi75}. But various
questions about the   distribution   of  ($d_{5/2}-p_{3/2}^{-1}$)
strengths among these states ensue, as does the question of how much
isospin mixing occurs. 

Particle transfer reactions, e.g. ${}^{17}$O($d,t$) and ${}^{17}$O($d, 
{}^3$He) to these 4$^-$ states give some useful indication of  the 
($d_{5/2}-p_{3/2}^{-1}$)    strength   distributions~\cite{Ma78}.
Assuming  that  the  reactions  occur by direct population of   the
$1p-1h$ components of the residual states,          the 
spectroscopic factors ($C^2S$) in the pure isospin limit must 
satisfy sum  rules, namely
%%%%%%%%%%%%%%%%%
\begin{equation}
  \sum (C^2S)_{(d,t)} = C^2S_{(d,{}^3{\rm He})}\ ,    
\end{equation}
%%%%%%%%%%%%%%%%%
and
%%%%%%%%%%%%%%%%%%
\begin{equation}
(C^2S)_{(d,t)}\vert_{(T=1)} = 
\frac{1}{2} C^2S_{(d,{}^3{\rm He})}\ . 
\end{equation}
%%%%%%%%%%%%%%%%%%%
 These relationships are satisfied approximately by experiment, and 
the extracted values for each transition suggest that the isovector
(18.977)  state  contains  97\%  of  the  isovector   particle-hole 
strength while the summed result of excitation of  the  (isoscalar) 
17.775 and  19.808  MeV  states  would  account  for  92\%  of that 
strength.        But extracted spectroscopic factors are very model 
dependent. Such percentages should be considered indicative at best.

\subsection
{A particle-hole model of the 4$^-$ excitations in ${}^{16}$O}

  Denoting the ground state of ${}^{16}$O by $\left|J^\pi;T\right\rangle = 
\left|0^+;0\right\rangle$, a simple model for the 4$^-$;$T_f$ states is
%%%%%%%%%%%%%%%%%%%
\begin{equation}
\left|4^-;T_f\right\rangle = 
\left|(p_{\frac{3}{2}}^{-1}d_{\frac{5}{2}})\,4^-;T_f\right\rangle = 
{\cal N} \left[ a^\dagger_{\left(\frac{5}{2},\frac{1}{2}\right)}
\times {\tilde a}^{\phantom{\dagger}}_{\left(\frac{3}{2},\frac{1}{2}\right)}  
\right]^{(4,T_f)}_{(M_4,M_{T_f=0})} \left|0^+;0\right\rangle\ ,
\end{equation}
%%%%%%%%%%%%%%%%%
in  which  the  particle-hole  operator  has  been  coupled in both
angular   momentum  and   isospin;  the  brackets  surrounding  the 
subscripts on the creation and annihilation operators indicate that 
full coupling.  This operator is defined by
%%%%%%%%%%%%%%%%%%
\begin{eqnarray}                  
\left[ a^\dagger_{\left(\frac{5}{2},\frac{1}{2}\right)}
\times {\tilde a}^{\phantom{\dagger}}_{\left(\frac{3}{2},\frac{1}{2}\right)}  
\right]^{(4,T_f)}_{(M_4, 0)} &&= \sum 
(-)^{\left( \frac{3}{2} - m_3\right)}
\left\langle \frac{3}{2}\, \frac{5}{2}\, m_3\, -m_5 \Bigl| 4\, -M_4
\right\rangle
\nonumber\\
&&  (-)^{\left( \frac{1}{2} - \alpha\right)}
\left\langle \frac{1}{2}\, \frac{1}{2}\, \alpha\, -\alpha \Bigl| T_f\, 0
\right\rangle
\ a^\dagger_{\frac{5}{2} m_5 \alpha} 
a^{\phantom{\dagger}}_{\frac{3}{2} m_3 \alpha}\ .
\end{eqnarray}
%%%%%%%%%%%%%%%%%%%%
      The summation is taken over all component particle projection 
quantum numbers, and $M_T = 0$ as we consider an $N = Z$   nucleus.
Herein we do  not consider  any reverse  amplitudes, i.e. 
$(p_{\frac{3}{2}}-d_{\frac{5}{2}}^{-1})$.     They will have very 
much smaller probability amplitudes,   and  none  if  the $d_{5/2}$ 
orbit in the ground state is vacant.   The normalization of both of 
these   states   then   is  fixed  by  the  fractional  occupancies 
$\sigma_{j}$ of nucleons of either type in the orbits ($j$) in  the
ground state.                    The normalization, as developed in 
Appendix~\ref{App-norm}, is
%%%%%%%%%%%%%%%%%%%%%%%
\begin{equation}
{\cal N} =  
\left[\left\{1 - \sigma_{\frac{5}{2}} \right\} 
\sigma_{\frac{3}{2}} \right]^{-\frac{1}{2}}\ .
\end{equation}
%%%%%%%%%%%%%%%%%%%%%%%

The same procedure allows a model of an 
IAS  in  ${}^{16}$F  though  an  additional  isospin   projection 
changing operator need be included.

     For inelastic scattering (or charge exchange leading to an IAS 
state in  ${}^{16}$F)  involving scattering from a nucleon in orbit 
$j_1$ with isospin $y$  and  then leaving  a nucleon in orbit $j_2$ 
with isospin $x$, the one body density matrix elements (OBDME), 
%%%%%%%%%%%%%%%%%%%%%%%%
\begin{equation}
S_{j_1j_2I}^{(x,y)} = \left< 4^-;T_f \left\| 
\left[a^\dagger_{j_2 x} \otimes {\tilde a}^{\phantom{\dagger}}_{j_1 y}
\right]^{(I)} \ \right\| 0^+;0\right>\ ,
\end{equation}
%%%%%%%%%%%%%%%%%%%%%%%%%%
are required.    Note that, in this case,  only angular momentum is 
coupled in the specification of the operator. That is designated by
the absence of brackets around the subscripts of the creation   and
annihilation operators and the use of $\otimes$ as coupling operator.     
Clearly the transfer quantum number is 
$I = 4$ and the combinations of  proton and  neutron  spectroscopic
amplitudes must effect an isospin change equal to $T_f$.

For inelastic scattering to the 4$^-$ states in ${}^{16}$O then, as
developed in Appendix~\ref{inel-Samps},
%%%%%%%%%%%%%%%%%%%%%%%%
\begin{equation}                  
S_{ \frac{3}{2}\frac{5}{2}4}^{(x)} 
= \frac{3}{\sqrt 2} \left\{ \delta_{(T_f 0)} + \delta_{(T_f 1)}
(-)^{\left( \frac{1}{2} - x\right)}\right\}\ \sqrt 
{\sigma_{\frac{3}{2}} \left[ 1 - \sigma_{\frac{5}{2}}\right] }\ , 
\label{specamps}
\end{equation}
%%%%%%%%%%%%%%%%%%%%
while for the charge exchange  ($p,n$)  reaction to the  IAS  $4^-$
state in ${}^{16}$F, as the struck nucleon must be  a  neutron   to
effect the charge transfer, $y = \frac{1}{2} = -x$.    Then, and as
developed in Appendix~\ref{pn-Samps},
%%%%%%%%%%%%%%%%%%%%%
\begin{equation}
S_{ \frac{3}{2}\frac{5}{2}4}^{(-\frac{1}{2},\frac{1}{2})} =
\delta_{(T_f 1)}\ 3  \sqrt{
\sigma_{\frac{3}{2}} \left[ 1 - \sigma_{\frac{5}{2}}\right]}\ . 
\end{equation}
%%%%%%%%%%%%%%%%%%%%%%
                 
\subsection{Configuration and isospin mixing} 

    We presume that the ground state of ${}^{16}$O has good isospin 
($T = 0$), but  configuration  mixing  will  certainly  reduce  the
$p_{3/2}$ occupancy from the closed shell value (of 4). A large basis
shell model calculation (in a complete $(0+2)\hbar\omega$  space) 
has been made of the ${}^{16}$O ground state~\cite{Ka96} and that 
gave values for $(2j + 1) \sigma_j$ respectively of 3.87 and 
0.14 for the $0p_{3/2}$  and $0d_{5/2}$  shells. To
describe the $4^-$ states  however,  we  consider  a  simple  three 
component basis space.    The three states of that basis the a pure 
particle-hole isovector states and two isoscalar states,
%%%%%%%%%%%%%%%%%%%
\begin{eqnarray}
\left|a\right> &=& \left|4^-; T=1\right>\  
= \left|(d_{\frac{5}{2}}p^{-1}_{\frac{3}{2}});\, 4^-;T=1\right>\ ,
\nonumber\\
\left|b\right> &=& \left|4^-; T=0\right>\ 
= \left|(d_{\frac{5}{2}}p^{-1}_{\frac{3}{2}});\, 4^-;T=0\right>\ ,
\nonumber\\
\left|c\right> &=& \left|(np - nh)\; 4^-; T=0\right>\ .
\label{basis}
\end{eqnarray}
%%%%%%%%%%%%%%%%%%%
With  these  it  is   possible   to  create  states   having   both 
configuration and isospin mixing.     In particular we suppose with 
this basis that there is an  optimal  prescription  for  the  three 
states of interest in ${}^{16}$O namely,
%%%%%%%%%%%%%%%%%%
\begin{equation}
\left(
\begin{array}{c}
\left|4^-; E_1 = 17.775\right>\\
\left|4^-; E_2 = 18.977\right>\\
\left|4^-; E_3 = 19.808\right>
\end{array}
\right) = \left(
\begin{array}{ccc} 
C_1(E_1) & C_0(E_1) & C_{np-nh}(E_1) \\
C_1(E_2) & C_0(E_2) & C_{np-nh}(E_2) \\
C_1(E_3) & C_0(E_3) & C_{np-nh}(E_3) 
\end{array}
\right)\;
\left(
\begin{array}{c}
\left|a\right>\\
\left|b\right>\\
\left|c\right>
\end{array}\right)\ .
\label{actuals}
\end{equation}
%%%%%%%%%%%%%%%%%
 We presume that the basis state $\left<c\right|$ cannot be excited 
by a simple 
particle-hole operator acting upon the ground state,    so that the
scattering amplitudes for the excitation of each state, $E_i$, then 
are weighted sums 
%%%%%%%%%%%%%%%%%%%%
\begin{equation}
{\cal M}(E_i) = C_0(E_i) {\cal M}_{\left<b\right|}  + 
C_1(E_i) {\cal M}_{\left<a\right|}\ , 
\end{equation}
%%%%%%%%%%%%%%%%%%%%%
where  the matrix  elements   ${\cal M}_{\left<x\right|}$    are    transition 
amplitudes    for    the    excitation    of    the   basis  states 
$\left\langle a \right|\, (T=1)$ and $\left\langle b \right|\, (T=0)$.   
However, 
it is also convenient to express the ${\cal M}(E_i)$ in a form
%%%%%%%%%%%%%%%%% 
\begin{eqnarray}
{\cal M}(E_1) &=& N_1\ \left\{ \cos(\epsilon_1)\, {\cal M}_{T=0}
+ \sin(\epsilon_1)\, {\cal M}_{T=1}\right\}
\nonumber\\
{\cal M}(E_2) &=& N_2\ \left\{ \cos(\epsilon_2)\, {\cal M}_{T=1}
+ \sin(\epsilon_2)\, {\cal M}_{T=0} \right\}
\nonumber\\
{\cal M}(E_3) &=& N_3\ \left\{ \cos(\epsilon_3)\, {\cal M}_{T=0}
+ \sin(\epsilon_3)\, {\cal M}_{T=1} \right\}\ ,
\label{mixediso}
\end{eqnarray}
%%%%%%%%%%%%%%%
where $N = \sqrt{C_0^2 + C_1^2}$ and $\cos(\epsilon) = C_T/N$. The 
coupling angle for each  state $\epsilon_i$, and the scales $N_i$, 
are  to  be  determined from data magnitudes of the measured cross 
sections.  For evaluations however,  individual proton and neutron 
amplitudes will be formed whence 
%%%%%%%%%%%%%%%%% 
\begin{eqnarray}
{\cal M}(E_1) &=& N_1\ \left\{ 
\left[\cos(\epsilon_1) + \sin(\epsilon_1)\right]\, {\cal M}_{\pi}
+ \left[\cos(\epsilon_1) - \sin(\epsilon_1)\right]\, 
{\cal M}_{\nu}\right\}
\nonumber\\
{\cal M}(E_2) &=& N_2\ \left\{ 
\left[\sin(\epsilon_2) - \cos(\epsilon_2)   \right]\, {\cal M}_{\pi}
+ \left[\sin(\epsilon_2) + \cos(\epsilon_2) \right]\, 
{\cal M}_{\nu} \right\}
\nonumber\\
{\cal M}(E_3) &=& N_3\ \left\{ 
\left[ \cos(\epsilon_3) + \sin(\epsilon_3) \right]\, {\cal M}_{\pi}
+ \left[\cos(\epsilon_3) - \sin(\epsilon_3)\right]\, 
{\cal M}_{\nu} \right\}\ .
\end{eqnarray}
%%%%%%%%%%%%%%%

%%%%%%%%%%  ELECTRON/PION/PROTON SCATTERING DETAILS %%%%%%%%%%%%
\section{Details of the scattering data analyses}
\label{Theory}

A particle-hole structure model for three $4^-$ states in ${}^{16}$O 
has been used in the past~\cite{Ba81,Hy87,Cl88} to analyze transverse 
magnetic $M4$ form factors extracted from 
inelastic electron  scattering  measurements~\cite{Mi75,Hy87}. So
also  have been  the
differential cross sections  from  the  inelastic   scattering  of
164~MeV pions~\cite{Ho80,Ba81} leading to those same three states. 
But the  inelastic scattering (of protons) to those states  has been 
considered    only   with   an   old    
phenomenological approach~\cite{Ba81,Pe83}.   We reconsider all of 
the data but now with  the  proton  inelastic  scattering  studied 
using a $g$-folding model of the scattering and with a large basis 
shell model wave function describing the ground state.  Details of 
the methods used in our analysis are given next in brief.

\subsection{Pion scattering}

The cross sections for the inelastic scattering of pions $\pi^\pm$
were evaluated using a conventional,  phenomenological,  distorted 
wave impulse approximation (DWIA).        The distorted waves were 
obtained using the optical model potential~\cite{St79},
%%%%%%%%%%%%%%%%%%%%%%%%
\begin{equation}
V^{(\pm)}(r) = -Z \frac{ \mu}{\hbar^2 } b_{0p} \rho_p(r)
 - N \frac{ \mu}{\hbar^2 } b_{0n} \rho_n(r)
 + Z b_{1p} \left\{ {\vec \nabla} \cdot \rho_p(r){\vec \nabla} 
\right\}    
 + N b_{1n} \left\{ {\vec \nabla} \cdot \rho_n(r){\vec \nabla} 
\right\}\ ,    
 \end{equation}
%%%%%%%%%%%%%%%%%%%%%%%
where the density function form is
%%%%%%%%%%%%%%%%%%
\begin{equation}
\rho_i(r) = \frac{2}{Z} \frac{1}{({a_i\sqrt \pi})^3}
\left[ 1 + \frac{Z-2}{3} \left(\frac{r}{a_i} \right)^2 
\exp\left\{-\left(\frac{r}{a_i} \right)^2 \right\}
\right]\ .
\end{equation}
%%%%%%%%%%%%%%%
with parameter values listed in Table~\ref{pitable}.
%%%%%%%%%%%%%%%%%%%%%%%%%%%%%%%%%%%%%%%%%%%%%%%%%%%%%%
\begin{table}[ht]
\begin{ruledtabular}
\caption{The parameter values of the $\pi^\pm$ optical potentials}
\label{pitable}
\begin{tabular}{ccc}
            & $\pi^+$ & $\pi^-$\\
\hline
$a_p = a_n$ & 1.805 & 1.805 \\
$b_{0p}$ & $-1.81 + 0.41\,i$& $\;\ 0.41 + 0.32\,i$ \\
$b_{0n}$ & $\;\ 0.41 + 0.32\,i$ & $-1.81 + 0.40\,i$ \\
$b_{1p}$ & $\;\ 5.70 + 14.0\,i$& $\;\ 1.53 + 4.69\,i$\\
$b_{1n}$ & $\;\ 1.53 + 4.69\,i$ & $\;\ 5.70 + 14.0\,i$\\
\end{tabular}
\end{ruledtabular}
\end{table}
%%%%%%%%%%%%%%%%%%%%%%%%%%%%%%%%%%%%%%%%%%%%%%%%%%%%%%
The inelastic scattering amplitudes also require specification  of 
the in-medium interaction of pions with bound nucleons as can   be
developed from a direct reaction scattering theory.     But as the 
$\pi N$ $g$-matrices in nuclear matter are not well  established
and, as the pion energy should give $\Delta$ dominance,  a simple 
contact form for the free $\pi N$ interaction    (the Kisslinger 
interaction)  has  been  assumed  as  the nuclear state transition 
operator to effect the   excitation  of  unnatural  parity states.    
Specifically, we have used
%%%%%%%%%%%%%%%%%%%%
\begin{equation}
t_{\pi N} = 
{\cal G}_0 \left[ {\vec \sigma}\cdot \left({\vec \kappa}
\times {\vec \kappa^\prime}\right) \right]\ 
\delta({\vec r}_\pi - {\vec r}_n)\ .
\end{equation}
%%%%%%%%%%%%%%%%%%%
The strength was taken by a match to that for the 18.977 MeV state 
excitation  assuming that it is a pure $T=1$ state as the particle 
transfer data suggest is most probable.  That such is adequate for
the interaction  will  be  confirmed  by  the results we find from 
simultaneous analysis of the scattering of $\pi^+$ and of  $\pi^-$ 
leading to the $4^-$ state excitation.     These details are gross 
simplifications of specifics of the scattering process but suffice
for use in assaying the amount of isospin mixing there is  in  the 
$4^-$ states. In the DWIA calculations, the single particle states
have been represented by harmonic oscillator wave functions.

\subsection{Electron scattering}

 Electron scattering form factor calculations have been made using
the  usual  point-particle,  one-body  current operators corrected 
for finite particle size and target recoil.           Relativistic 
kinematics were used to specify the momentum transfer values.  The 
single particle bound states used in our first analyses were 
harmonic oscillators for an oscillator length of 1.7 fm.  
Subsequently Woods-Saxon (WS) bound state wave functions were used,
with which  other studies~\cite{Cl88} have found improved shapes
of calculated form factors. 
Thus we have used WS functions for bound nucleon states
with binding energies (of the dominant orbits and in MeV) being
$-$38.0 ($0s_{1/2}$), $-$14.5 ($0p_{3/2}$), $-$10.5 ($0p_{1/2}$),
$-$4.28 ($0d_{5/2}$),  $-$3.87 ($1s_{1/2}$),  and $-$ 3.59 ($0d_{3/2}$). 
Higher shell (small occupancy)  orbits  were included and  defined with 
binding energies between $-1.5$ and $-1.0$ MeV; their exact values are 
not very 
important for they have little impact on results.

\begin{figure}
\scalebox{0.65}{\includegraphics*{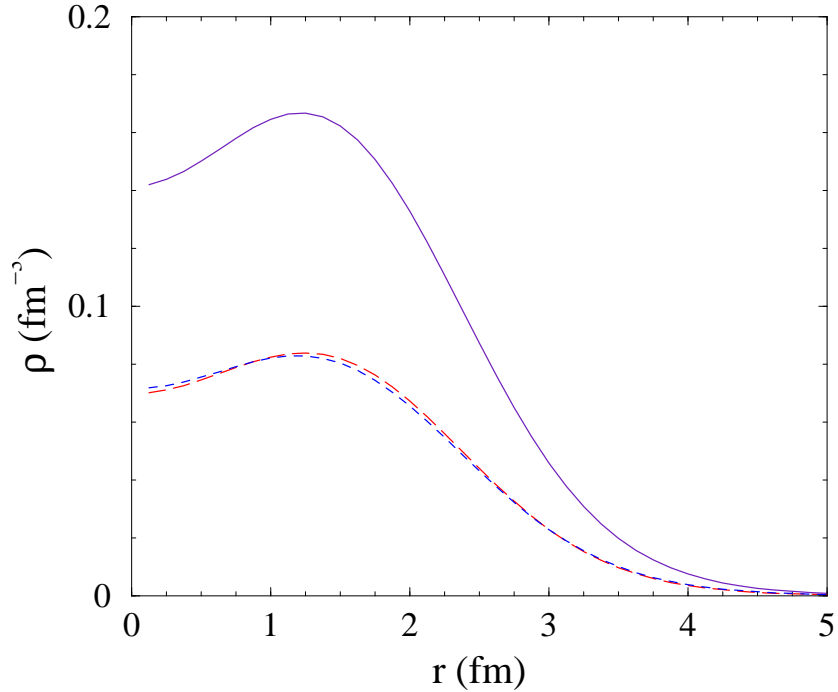}}
\caption{\label{Fig1-WS-dens}(Color online)
Proton  (large dash),  neutron (small dash),  and  total   (solid) 
mass densities  for  ${}^{16}$O  as  defined  by the WS  bound 
state functions.}
\end{figure}

Analyses of longitudinal form factors from elastic scattering of 
electrons, as
well as of cross sections and spin measurable from the elastic scattering
of protons, when the nucleus is treated microscopically, require the 
OBDME for the target ground state. Frequently they are just the nucleon
shell occupancies in the ground state. A complete $(0+2)\hbar\omega$ 
shell model calculation of ${}^{16}$O~\cite{Ka96}, gave the the OBDME 
numbers (the dominant values) listed in Table~\ref{SMoccups}.  The 
first two columns are  
the shell occupancies while the cross-shell values are given in the 
last three columns.
%%%%%%%%%%%%%%%%%%%%%%%%%%%%%%%%%%%%%%%%%%%%%%%%%%%%%%%% 
\begin{table}[ht]
\begin{ruledtabular}
\caption{The dominant shell occupancies and cross-shell terms for 
elastic scattering}
\label{SMoccups}
\begin{tabular}{cccccc}
$(nl_j)_1$ & occupancy &  $(nl_j)_1$ & $(nl_j)_2$ & 
$\langle \Psi_{gs} \left\| \left[a_{j_1}^\dagger 
\otimes {\tilde a}_{j_2}\right]^0 \right\| \Psi_{gs}\rangle$\\
\hline
 $0s_{1/2}$ & 1.9993 & $0s_{1/2}$ & $1s_{1/2}$ & 0.0334\\
 $0p_{3/2}$ & 3.8706 & $0p_{3/2}$ & $1p_{3/2}$ & 0.2036\\
 $0p_{1/2}$ & 1.8942 & $0p_{1/2}$ & $1p_{1/2}$ & 0.0587\\
 $0d_{5/2}$ & 0.1413 & & & \\
 $0d_{3/2}$ & 0.0678 & & & \\
 $1s_{1/2}$ & 0.0107 & & & \\
\end{tabular}
\end{ruledtabular}
\end{table}
%%%%%%%%%%%%%%%%%%%%%%%%%%%%%%%%%%%%%%%%%%%%%%%%%%%%%%%%%%%%%
When weighted by the ground state OBDME listed in Table~\ref{SMoccups},
the WS functions gave the densities that are shown in  
Fig.~\ref{Fig1-WS-dens}. Those densities have 
root mean square radii of 2.61, 2.65, and 2.63 fm  for  the proton, 
neutron, and total mass densities respectively; in good agreement with
the value of 2.62 fm~\cite{Mi00}, and like that, a bit smaller than the
experimental value of $2.712\pm 0.015$ fm. 
%%%%%%%%%%%%%%%%%%%%%%%%%%%%%%%%%%%%
%%%%%%%%%%%%%%%%%%%%%%%%%%%%%%%%%%%%%%

Electron scattering data (form factors) for scattering on ${}^{16}$O
have been measured; the elastic form factors by Sick and 
McCarthy~\cite{Si70}  and the relevant 
$M4$ form factors by  Hyde-Wright {\it et al.}~\cite{Hy87}. 
While the elastic data has been 
measured to momentum transfers up 4 fm$^{-1}$, the $M4$ data span
an effective 
momentum transfer range from 0.8 to 2.6 fm$^{-1}$. Of the three states 
considered, form factors from excitation of the        (dominantly 
isoscalar) 17.775 MeV state and from the    (dominantly isovector) 
18.877 MeV state have been well resolved.   Only statistical upper 
bounds are known for the form factor from excitation of the 19.808 
MeV state.

Finally note that we have not included any MEC corrections in our 
calculations of the $M4$ form factors. They have been considered in 
the past~\cite{Cl88} and found to be relatively minor in effect. 
Therefore MEC should not alter whatever general description of structure 
we can deduce from comparison of evaluated form factors   with use 
of the same structures in the analyses of other scattering data.

\subsection{Proton inelastic and charge exchange scattering}

 Cross sections for inelastic proton scattering exciting the $4^-$ 
states in ${}^{16}$O, and for the charge exchange  reaction to the
$4^-$ states in ${}^{16}$F,  have  been  evaluated  using  a fully 
microscopic DWA theory of the processes~\cite{Am00}. The distorted
waves are generated from optical potentials formed  by folding  an
effective in-medium $NN$ interaction set for each energy  from the
BBG $g$-matrices with the OBDME of the target  states~\cite{Am00}. 
In coordinate space that effective $NN$ interaction is a mix 
of central, two-body spin-orbit and tensor forces all having  form 
factors that are sums of Yukawa functions.     Then with the Pauli
principle taken into account,   optical potentials from the folding
are complex, nonlocal, and energy dependent.   Such are formed and
used in the DWBA98 program~\cite{Ra98} to predict elastic 
scattering observables. That same program finds distorted wave functions 
from those potentials for use in  DWA  calculations  of  the inelastic 
scattering and  
charge exchange cross sections.      The transition amplitudes for 
nucleon   inelastic   scattering    from    a    nuclear    target 
$J_i \longrightarrow J_f$ have the form~\cite{Am00}
%%%%%%%%%%%%%%%%%%%%%%%%%%%%%%%%%%%%%%%%%%%%%%%%
\begin{eqnarray}
{\cal T} &=& T^{M_fM_i\nu^\prime\nu}_{J_fJ_i}(\Omega_{sc})
\nonumber\\
&=&\left\langle \chi^{(-)}_{\nu^\prime}({\bf k}_o0)\right|
\left\langle\Psi_{J_fM_f}(1 \cdots A) \right|
\; A{\bf g}_{eff}(0,1)\;
 {\cal A}_{01} \left\{ \left| \chi^{(+)}_\nu ({\bf
k}_i0) \right\rangle \right.  \left. \left| \Psi_{J_iM_i}
 (1\cdots A) \right\rangle \right\}\ ,
\end{eqnarray}
%%%%%%%%%%%%%%%%%%%%%%%%%%%%%%%%%%%%%%%%%%%%%%%%%
where $\Omega_{sc}$ is the scattering angle and  ${\cal A}_{01}$ is 
the antisymmetrization operator. Then a cofactor expansion of   the
target states, 
%%%%%%%%%%%%%%%%%%%%%%%%%%%%%%%%%%%%%%%%%%%%%%%%%%%%%%%%%%%%%%
\begin{equation}
\left| \Psi_{JM}(1,\cdots A) \right\rangle = \frac{1}{\sqrt{A}} \sum_{j,m}
\left| \varphi_{jm}(1) \right\rangle\,
a_{jm}(1)\, \left| \Psi_{JM}(1,\cdots A) \right\rangle\ ,
\label{cofactor}
\end{equation}
%%%%%%%%%%%%%%%%%%%%%%%%%%%%%%%%%%%%%%%%%%%%%%%%%%%%%%%%%%%%%%%%
allows  expansion of the  scattering  amplitudes in the form  of 
weighted two-nucleon elements since the terms 
$a_{jm}(1)\, \left| \Psi_{JM}(1,\cdots A)\right\rangle$ in 
Eq.~\ref{cofactor} are independent of coordinate `1'. Thus 
%%%%%%%%%%%%%%%%%%%%%%%%%%%%%%%%%%%%%%%%%%%%%%%%%%%%%%%%%%%%%%%%
\begin{eqnarray}
{\cal T} &=& \sum_{j_1,j_2} \left\langle\Psi_{J_fM_f}(1,\cdots A) \right|
a^{\dagger}_{j_2m_2}(1) a_{j_1m_1}(1) \left| \Psi_{J_iM_i}(1,\cdots A) 
\right\rangle 
\nonumber\\
&&\hspace*{1.0cm} \times
\left\langle \chi^{(-)}_{\nu^\prime}({\bf k}_o0)\right|
\left\langle \varphi_{j_2m_2}(1) \right| \ {\bf g}_{eff}(0,1)\ 
{\cal A}_{01} \left\{ \left| \chi^{(+)}_\nu ({\bf k}_i0) 
\right\rangle
\ \left| \varphi_{j_1m_1} (1) \right\rangle \right\}
\nonumber\\
&=& \sum_{j_1,j_2,m_1,m_2,I(N)}
(-)^{(j_1-m_1)} \frac{1}{\sqrt{2J_f+1}}\,  
\left< J_i\, I\, M_i\, N \vert J_f\, M_f \right>
\left< j_1\, j_2\, m_1\, -m_2 \vert I\, -N \right>
\, S_{j_1\, j_2\, I}^{(J_i \to J_f)}\ 
\nonumber\\
&&\hspace*{2.0cm} \times
\Bigl< \chi^{(-)}_{\nu^\prime}({\bf k}_o0)\Bigr|
\left\langle \varphi_{j_2m_2}(1) \right| \
{\bf g}_{eff}(0,1)\ 
 {\cal A}_{01} \left\{ \left| \chi^{(+)}_\nu ({\bf k}_i0) 
\right\rangle
\ \left| \varphi_{j_1m_1} (1) \right\rangle \right\}\ ,
\end{eqnarray}
%%%%%%%%%%%%%%%%%%%%%%%%%%%%%%%%%%%%%%%%%%%%%%%%%%%%%%%%%%%%%%%%
where reduction of the structure factor to  OBDME  for    angular 
momentum transfer values $I$ follows that  developed  earlier  for 
excitation from ground of the $4^-$ states.

  The effective interactions $g_{eff}(0,1)$ used in the folding to
get  the  optical  potentials   have   also   been   used  as  the  
transition  operators  effecting  the excitations    (of the $4^-$ 
states). As with the generation of the elastic scattering   and so 
also the  distorted wave functions for use in the DWA evaluations, 
antisymmetry   of   the  projectile  with   the  individual  bound 
nucleons is treated exactly.   The associated knock-out (exchange) 
amplitudes   contribute   importantly   to  the  scattering  cross  
section, both in magnitude and shape~\cite{Am00}.

Thus only the structure details are left to be specified. They are
the    spectroscopic    amplitudes    as   have  been  defined  in  
Sec.~\ref{spectroscopy},   and   the   single nucleon bound  state 
wave functions.  Those are exactly the same as we have used in      
analyzing electron scattering form factors and as specified in 
detail in the previous subsection.

%%%%%%%%%%%%%%%%%%%   RESULTS %%%%%%%%%%%%%%%%%%%%%%%%%%%%%%%%%
\section{Results of calculations}
\label{Results}

   Analyses of the data from the scattering of pions 
leading to the set of $4^-$ states are considered first since they 
most strongly provide evidence of isospin mixing in  those states. 
Then, in the second subsection, we present the results of elastic
scattering of electrons and of protons from ${}^{16}$O. Thereafter
we give results of evaluations of the $M4$ form factors from 
electron scattering while in the last subsection we present 
results of excitations of the same 4$^-$ states by 
inelastic scattering and charge exchange scattering of protons.

\subsection{Pion scattering to the $4^-$ states in ${}^{16}$O}

The isospin mixing in the three states of  interest  is  indicated
most clearly by considering the  peak cross section values in pion
scattering.    Specifically we made calculations of $\pi^\pm$ peak 
height values and  plotted ratios of actual data values over those
theoretical estimates,  irrespective of whether those peaks  occur 
exactly at the same values of momentum transfer, and with whatever
overall strengths  $N_i$  were  required  to  match observed  peak 
values.        Our base scale choice for the pion-nucleon strength
${\cal G}_0$   was  that  which  best fit the actual data from the
excitation of the 18.977 MeV state.          The data from $\pi^+$ 
scattering essentially has the same magnitude as that from $\pi^-$ 
scattering~\cite{Ho80} and  that  near equality strongly suggested
that the state was purely isovector in nature~\cite{Ho80}.

%%%%%%%%%%%%%%%%%%%%%%%%%%%%%%%%%%%%%%%%%
\begin{figure}[ht]
\scalebox{0.7}{\includegraphics*{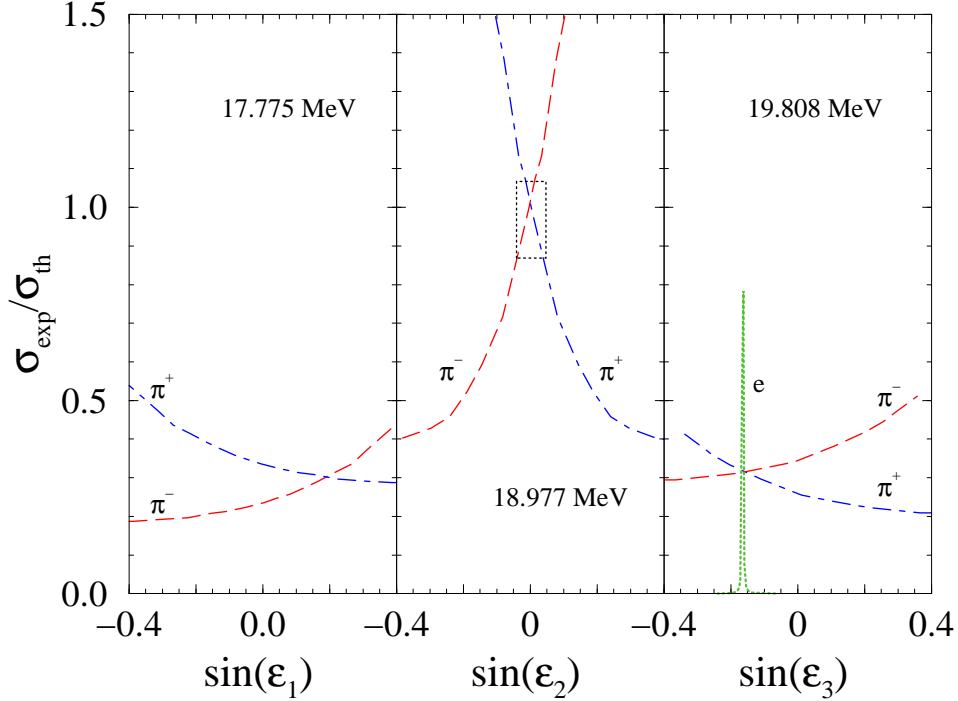}}
\caption{\label{Fig2-epi-ratios}(Color online)
Peak height ratios of experiment to theory for $\pi^\pm$ scattering 
to the three  $4^-$  states in  ${}^{16}$O as  functions of isospin
mixing.}
\end{figure}
%%%%%%%%%%%%%%%%%%%%%%%%%%%%%%%%%%%%%%%%

    With that interaction strength we then evaluated the $\pi^\pm$ 
cross sections from excitation of the other (dominantly isoscalar) 
$4^-$ states,   ascertaining  the normalization scales required to
fit the data and allowing variation in isospin mixing.         The 
variation of the peak height ratios $\sigma_{exp}/\sigma_{th}$  so 
obtained are shown in   Fig.~\ref{Fig2-epi-ratios} as functions of
the sines of the coupling angles $\epsilon_i$.  Such reflects  the
amount of  isospin  mixing  one might expect for each of the three
states. A variation of the peak cross section ratio found with the
scattering of electrons exciting  the  19.808  MeV  state is  also 
shown as it also emphasizes the isospin mixing expectation for that 
state. The variation with isospin mixing of the $\pi^+$ scattering
cross section peak ratio is markedly different to that for $\pi^-$ 
scattering in all three cases and those variations are quite sharp.
Where the two projectile variations cross then indicates the amount
of isospin mixing to be expected. With the 19.808 MeV  excitation, 
the electron scattering data are only known as upper bounds and so
using those (small) values gives a very sharp change when  isospin 
mixing in that dominantly isoscalar state is used to calculate the
relevant $M4$ transverse electric form factor.        That variation 
confirms well the suggested isospin admixing with $\sin(\epsilon_3)
= -0.17$.  Thus we assumed, initially as least, that the   isospin 
mixing   in   the   three   $4^-$   states   of   ${}^{16}$O to be 
$\sin(\epsilon_1) = 0.2$ for the 17.775~MeV state, 
$\sin(\epsilon_2) = 0$ for the 18.977~MeV state, and  
$\sin(\epsilon_3) = -0.17$ for the 19.808~MeV state.

Holtkamp~{\em et al.}~\cite{Ho80} quote a ratio of $\pi^\pm$ summed
yields for excitation of the  18.977 MeV state as $0.964 \pm 0.08$
so allowing some small amount of isospin mixing.    A conservative 
estimate is shown by the width of the dotted box around  the cross 
over point in the middle panel of Fig.~\ref{Fig2-epi-ratios}.

%%%%%%%%%%%%%%%%%%%%%%%%%%%%%%%%%%%%%%%%%
\begin{figure}
\scalebox{0.7}{\includegraphics*{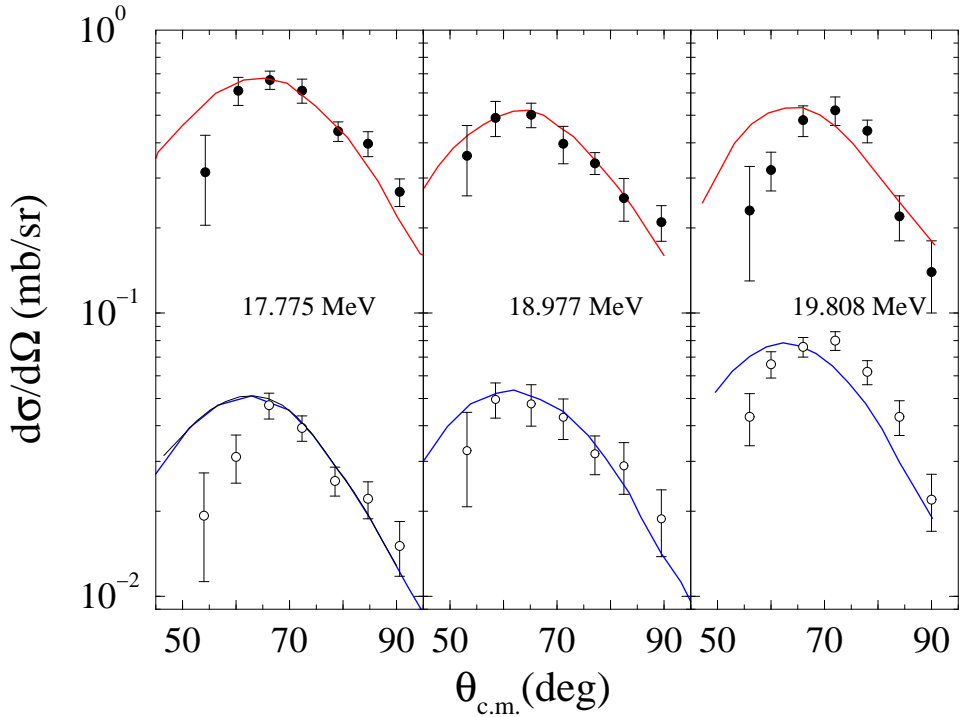}}
\caption{\label{Fig3-pixsec}(Color online)
Differential cross sections for $\pi^\pm$ scattering to the three  
$4^-$  states in  ${}^{16}$O with fits requiring overall scale 
factors as given in the text.}
\end{figure}
%%%%%%%%%%%%%%%%%%%%%%%%%%%%%%%%%%%%%%%%
        In Fig.~\ref{Fig3-pixsec}, the differential cross sections 
measured~\cite{Ho80}  for the inelastic scattering of  pions  from  
${}^{16}$O and leading  to the $4^-$ states are compared with DWIA
calculated results.   The interaction operator strength was set by  
assuming   the   18.977  MeV  state  to  be  described by the pure 
isovector basis state $\left| a \right>$ of Eq.~(\ref{basis}).     That is 
consistent also with previous DWIA analyses~\cite{Ca83} of the same
data  as  well  as  of  cross  sections  from  the  excitation  of 
``stretched'' $6^-$ states in ${}^{28}$Si. 
The fits to the data from the other states required scales of  0.3
and 0.34, for the 17.775 and 19.808~MeV states respectively   when 
the  isospin  admixing  suggested  by the peak height ratio  study 
described above.      That isospin mixing was particularly crucial 
since the relative peak height ratios of the  $\pi^+$ and  $\pi^-$ 
vary so markedly with the mixing. However, the shapes of the cross 
sections calculated for the dominantly  isoscalar  states  do  not
match  the data as well as they could.           Indeed a previous
analysis~\cite{Ca83} used spin transition densities extracted from 
electron  scattering which resulted in better cross sections shapes
than we show.        But such improvements are not central to this 
composite data study and,  in any event,  would  require a  better
specification   of   the   $\pi^\pm N$  $g$-matrices   including 
allowance   of   the   $T = \frac{1}{2}$   transition  amplitudes; 
components found to be of some import, despite $\Delta$ dominance,   in 
studies of $\pi^\pm$ asymmetries  in cross sections for excitation 
of   the   $\frac{9}{2}^+$   state   in    ${}^{13}$C~\cite{Am81}. 
Furthermore, any such refinement has not been made since there are 
also vagaries with the phenomenology involved in the calculations. 

\subsection{Electron and proton elastic scattering from ${}^{16}$O}

The charge density shown in Fig.~\ref{Fig1-WS-dens}, is tested by using 
it in evaluations of the 
longitudinal form factor for elastic electron scattering. 
%%%%%%%%%%%%%%%%%%%%%%%%%%%%%%%%%%%%
\begin{figure}[ht]
\scalebox{0.55}{\includegraphics*{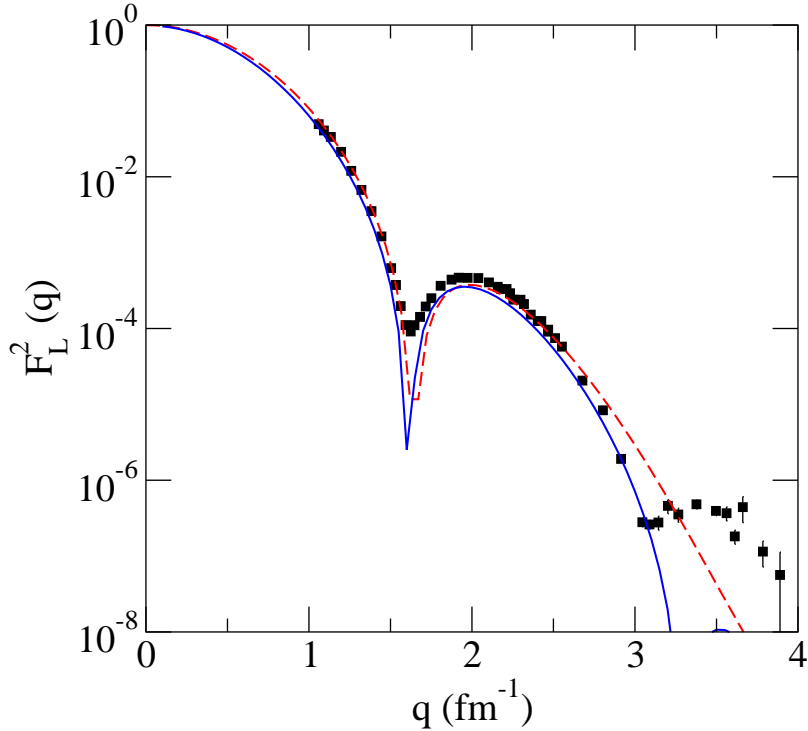}}
\caption{\label{Fig4-Elect-elastic}(Color online)
The longitudinal form factor from electron elastic scattering on 
${}^{16}$O. The data~\cite{Si70} are compared with the results of
calculations made using harmonic oscillator ($b = 1.7$ fm) wave functions
(dashed curve) and with the set of WS functions (solid curve) defined
in the text.}
\end{figure}
%%%%%%%%%%%%%%%%%%%%%%%%%%%%%%%%%%%%%%
The results
of using harmonic oscillators ($b = 1.7$ fm) and WS functions are compared
with the data~\cite{Si70} in Fig.~\ref{Fig4-Elect-elastic}. The data (filled 
squares) show that the oscillator (dashed curve) is a good result to 
$2.5$ fm$^{-1}$ while the WS result (solid curve does as well to the second 
minimum in data at 3 fm$^{-1}$.
We note that a much better fit to these data above 3 fm$^{-1}$ has been 
found
by Mihaila and Heisenberg~\cite{Mi00} by using an $\exp(S)$ 
coupled-cluster expansion approach. But to achieve those remarkable
results necessitated use of an effective 
50$\hbar\omega$ space, two-body currents, and Coulomb distortion; things 
not included in the simple approach we have taken since our purpose is
to correlate major effects in scattering of different probes
with analyses of data for which momentum transfer values are less than 
$\sim 3$ fm$^{-1}$.

A further test of the propriety of the structure assumed for the ground
state is to use that in analyses of the elastic scattering of protons  
from ${}^{16}$O.  Those we have made using the $g$-folding 
approach~\cite{Am00}  that  has  proved quite successful in recent
years.  Details of the approach are given in the review~\cite{Am00} as 
are details of the Melbourne effective $NN$ interaction that is actually
folded with the spectroscopy. 

\begin{figure}
\scalebox{0.65}{\includegraphics*{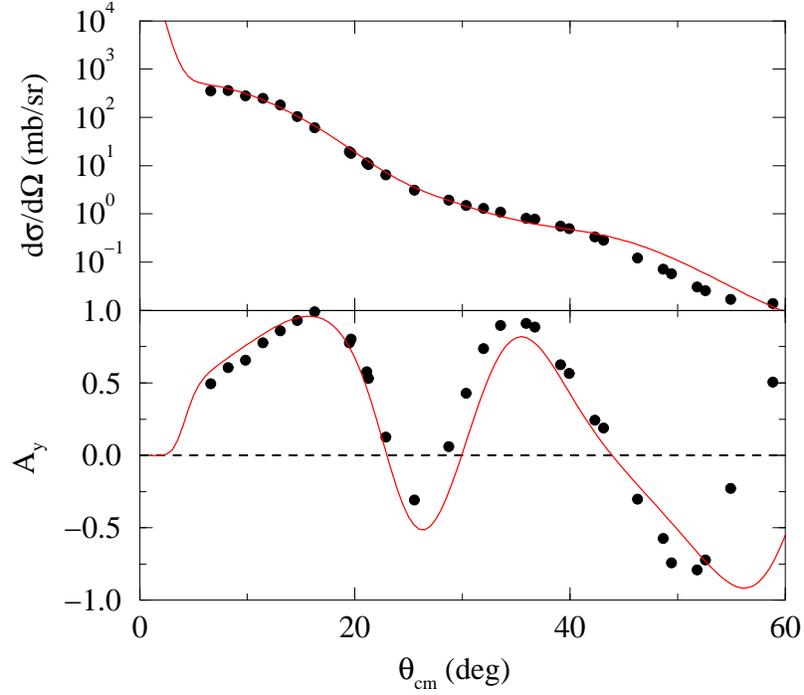}}
\caption{\label{Fig5-elas200}(Color online)
Differential Cross section (top) and analyzing power (bottom)
for the elastic scattering of 200 MeV protons from ${}^{16}$O.} 
\end{figure}

We consider proton scattering at 200 MeV first since past studies~\cite{Am00}
of data from many nuclei gave great confidence that
the effective $NN$ interaction at that energy was good, and that the method 
of analysis was appropriate.
In Fig.~\ref{Fig5-elas200}, 200 MeV data~\cite{Se93} from ${}^{16}$O (filled 
circles)  are compared with the results 
(solid curves) found using the $g$-folding method with the appropriate 
(200 MeV) effective $NN$ interaction, WS bound state functions, and 
(ground state) OBDME from the complete $(0+2)\hbar \omega$ shell model.
These predictions are in very good agreement with that data and especially
%%%%%%%%%%%%%%%%%%%%%%%%%%%%%%%%%%%%%%%%%%%%%%%%%%%%
%%%%%%%%%%%%%%%%%%%%%%%%%%%%%%%%%%%%%%
the good description of the analyzing power in particular suggests that the 
ground state spectroscopy we have chosen is realistic.  

\begin{figure}
\scalebox{0.65}{\includegraphics*{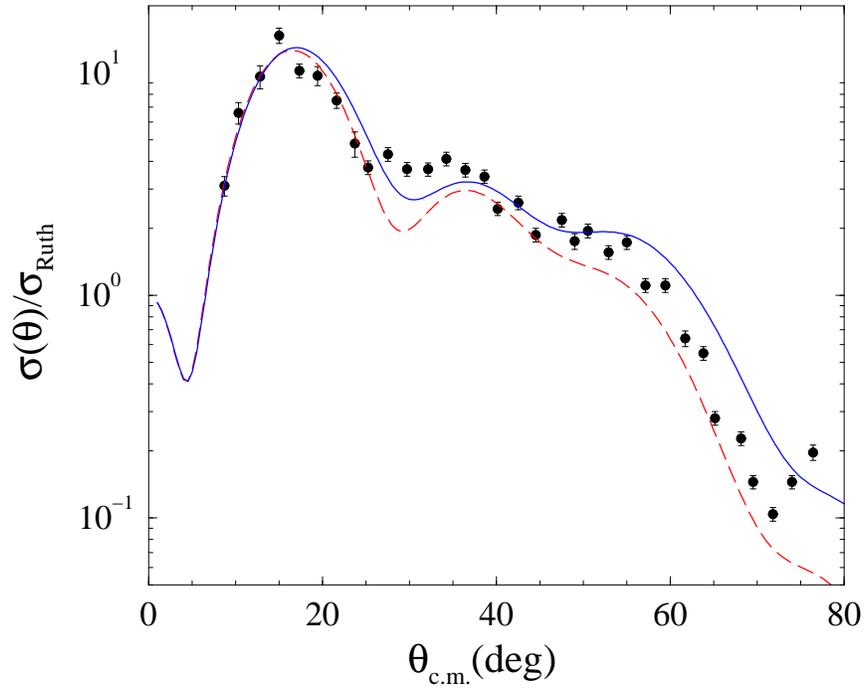}}
\caption{\label{Fig6-elas135}(Color online)
Ratio to  Rutherford  cross sections for the elastic scattering of 
135 MeV protons from ${}^{16}$O. The solid and dashed curves display the
results found using WS and oscillator bound state wave functions 
respectively.} 
\end{figure}

Using that spectroscopy and simply changing the effective interaction
to that appropriate for 135 MeV protons lead then to the results shown
in Fig.~\ref{Fig6-elas135}. Therein, the
elastic scattering cross sections of 135 MeV protons from    
${}^{16}$O as a ratio to Rutherford scattering  are   compared with  
data~\cite{Am84}. 
%%%%%%%%%%%%%%%%%%%%%%%%%%%%%%%%%%%%
%%%%%%%%%%%%%%%%%%%%%%%%%%%%%%%%%%%%%%
The WS folding result is displayed by the solid curve while the dashed 
curve portrays that found by using the oscillator functions instead. There
are some differences between these results but the WS result is preferred
since the cross section forward of 60$^\circ$ is the most significant as a 
test of the model.  

\subsection{$M4$ form factors from inelastic electron scattering}

%%%%%%%%%%%%%%%%%%%%%%%%%%%%%%%%%%%%%%%%%
\begin{figure}
\scalebox{0.7}{\includegraphics*{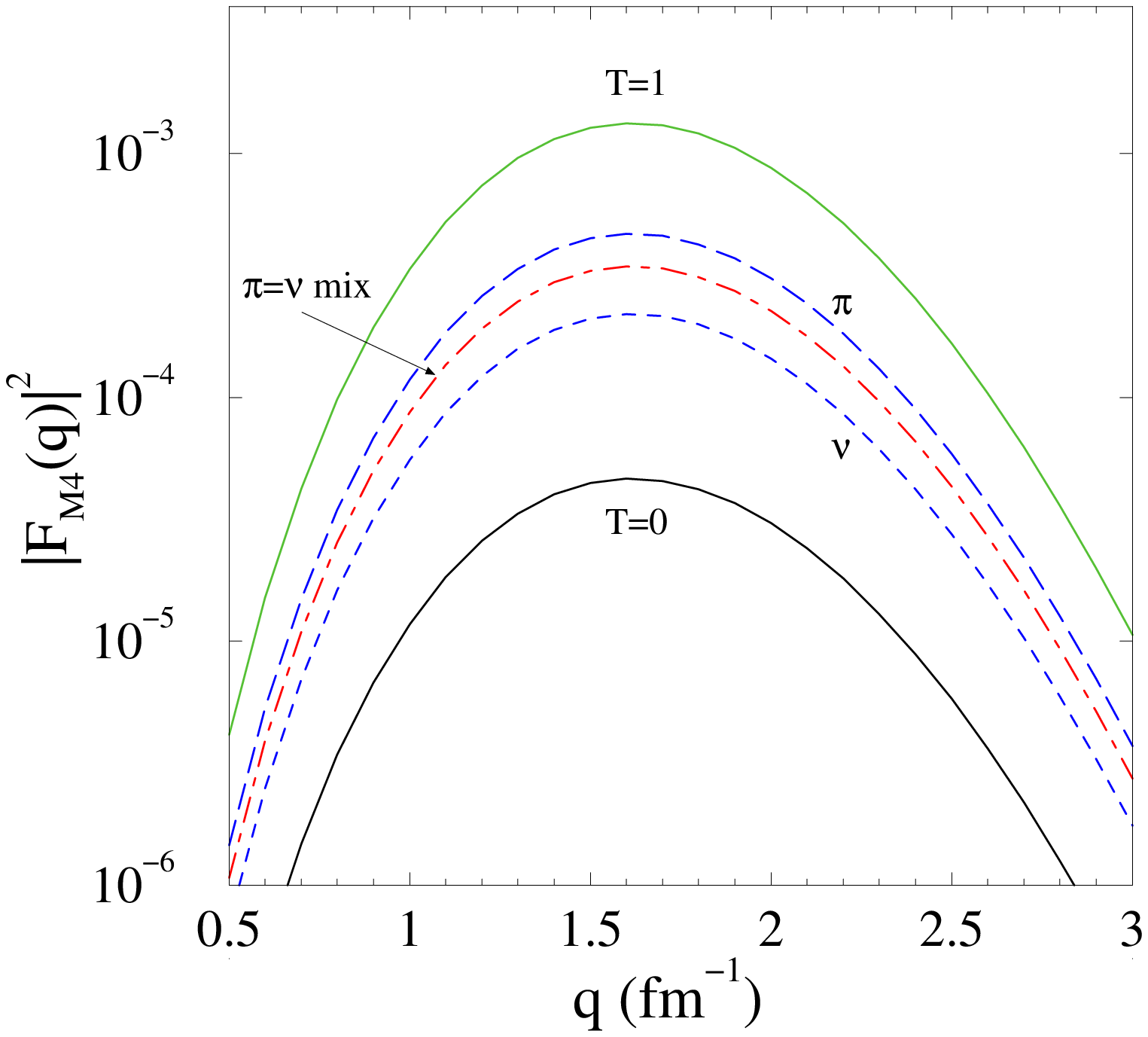}}
\caption{\label{Fig7-elect-comps}(Color online)
$M4$   form  factors  from  electron  scattering to the pure isospin 
``stretched'' states of Eq.~(\ref{basis}).}
\end{figure}
\begin{figure}
\scalebox{0.7}{\includegraphics*{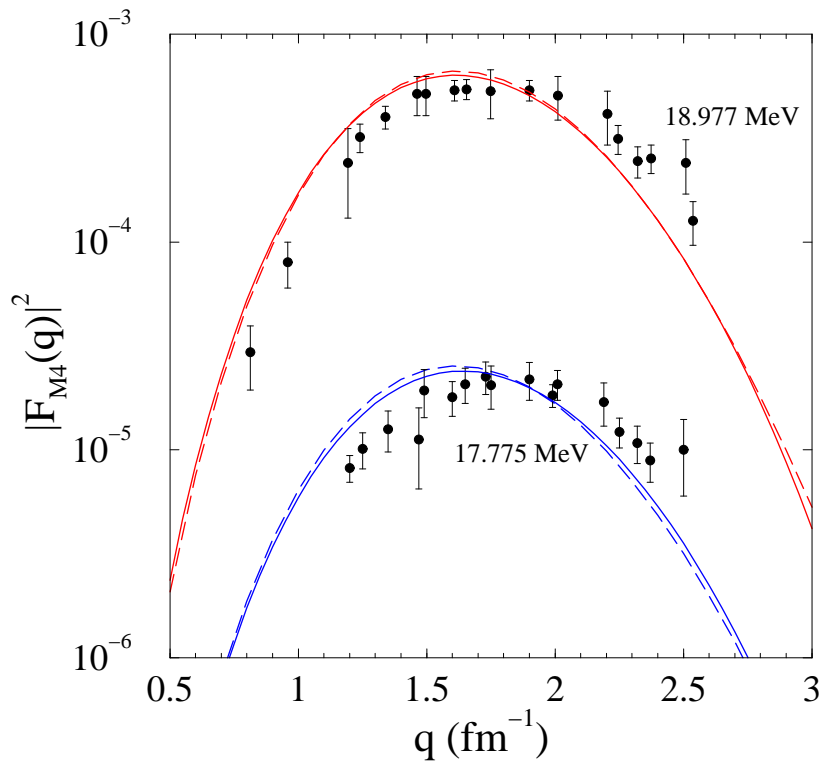}}
\caption{\label{Fig8-elect}(Color online)
$M4$ form factors from electron scattering to the 18.977~MeV and to 
the 17.775~MeV $4^-$  states in  ${}^{16}$O.} 
\end{figure}

%%%%%%%%%%%%%%%%%%%%%%%%%%%%%%%%%%%%%%%%%%%%%%%%%%%%%%%%%%
     The electron scattering data for excitation of the 17.775 and  
18.977~MeV states are well defined but only a weak upper limit  is 
known   for   the $M4$ form factor from excitation of the 19.808~MeV 
state.    As shown earlier, the latter quite strongly confirms the 
degree of isospin mixing in that state. First we consider the form 
factors and the separate proton and neutron components that  arise 
when excitations of the pure isospin basis states are considered. 
The results displayed in Fig.~\ref{Fig7-elect-comps} were generated 
using WS wave functions.
The  separate proton ($\pi$) and neutron ($\nu$) results are shown 
by the long and short dashed lines respectively.    The dot-dashed
line is the pure proton   (and neutron)  form factor  one finds by 
adjusting the weight values to get identical  proton  and  neutron 
amplitudes.  Clearly the 
underlying amplitudes interfere  constructively and  destructively 
respectively  to  yield the  isovector  and  isoscalar  excitation 
results.   By small weightings, magnitude equal proton and neutron 
amplitudes can be found and that defines the isospin mixing required
to give nett vanishing form factors under conditions of destructive 
interference.        That occurs essentially with the 
excitation of the 19.808~MeV state.   Likewise finite results for 
the $M4$ form factors from excitation of the other two actual states 
can be found by such mixing.     However, in those cases, there is 
also configuration mixing to be considered.       By selecting the 
isospin admixtures as suggested from the  pion scattering results, 
the $M4$ form factors from excitation of the  17.775 and  18.977~MeV 
states   by    electron    scattering    are    those  shown    in 
Fig.~\ref{Fig8-elect}.  The dashed curves are those obtained using 
oscillator functions  whilst those found on using WS functions 
are displayed by the solid curves.       Both sets of calculations 
required the scalings (down) from the values found assuming that
there was no spreading of particle-hole strength over other $4^-$
states in the spectrum.  The results are in reasonable 
agreement with those made previously~\cite{Hy87} in which oscillator
wave functions with oscillator lengths of 1.58 fm were used to improve 
the comparison of calculated results with the data. As we have used 
a larger oscillator length, our form factors are more sharply 
peaked but, as evident, they are consistent with results found
using the WS functions and with the elastic scattering data.  

  Specifically isospin mixing has been taken into account with the 
analysis  of  the  17.775~MeV  excitation  with  the mixing angle, 
$\sin(\epsilon_1) = 0.2$,   while  the  18.977~MeV transition  was 
taken as a pure isovector  excitation. 
%%%%%%%%%%%%%%%%%%%%%%%%%%%%%%%%%%%%%%%%%
%%%%%%%%%%%%%%%%%%%%%%%%%%%%%%%%%%%%%%%%
But  a  small isoscalar admixture in this does  not vary  the peak 
magnitude markedly.       Thus the pure isovector characterization 
suggested by the pion scattering is neither abrogated  nor totally 
supported  by these results.  However, they do identify an overall 
normalization for the state of $N_2 = 0.7$.     In contrast,  the 
calculated form factor for the (dominantly isoscalar)   17.775~MeV 
transition varies noticeably in magnitude  with  isospin admixture 
and using the amount selected from the $\pi^\pm$  scattering  data 
analyses   requires   a   normalization $N_1 = 0.36$ when harmonic 
oscillators are taken as the bound state functions. But the 
dominantly isoscalar transition result arises from an adjustment  
to the amount  of  destructive interference between the proton and 
neutron $M4$ amplitudes.  Such makes assessment of the configuration 
mixing effects difficult.  We place more reliability in the proton 
inelastic scattering results that we discuss next, especially since
the separate (bound) proton and neutron amplitudes   constructively 
interfere to define the cross sections.

%%%%%%%%%%%%%%%%%%%%%%%%%%%%%%%%%%%%%%%%%%%%%%%%%%%%%%%%%%
\subsection{Proton scattering leading to the $4^-$ states in 
${}^{16}$O}

In this subsection, we present the results of analyses of 135~MeV 
inelastic scattering data~\cite{He79}, of 135~MeV ($p,n$) data   (to
the   IAS   $4^-$  state in ${}^{16}$F) and of the analyzing power 
measured in that experiment~\cite{Ma82}.          The latter is of 
particular interest in that spin observables    usually  are quite
sensitive to details in the calculations. 
   As we shall see, at 135~MeV, the charge exchange data track the 
inelastic proton scattering cross section quite well  with  but  a 
scale factor of 2. That should be so if the state in ${}^{16}$F is 
a true IAS of the 18.977~MeV isovector state in ${}^{16}$O.

All calculations have been made using a DWA approach with the distorted
waves being those generated from evaluation of the
elastic scattering of 135~MeV  protons from ${}^{16}$O and the 
transition operator being the same 
effective interaction that we used in the $g$-folding solution of that
elastic  scattering.  
%%%%%%%%%%%%%%%%%%%%%%%%%%%%%%%%%%%%
\begin{figure}
\scalebox{0.65}{\includegraphics*{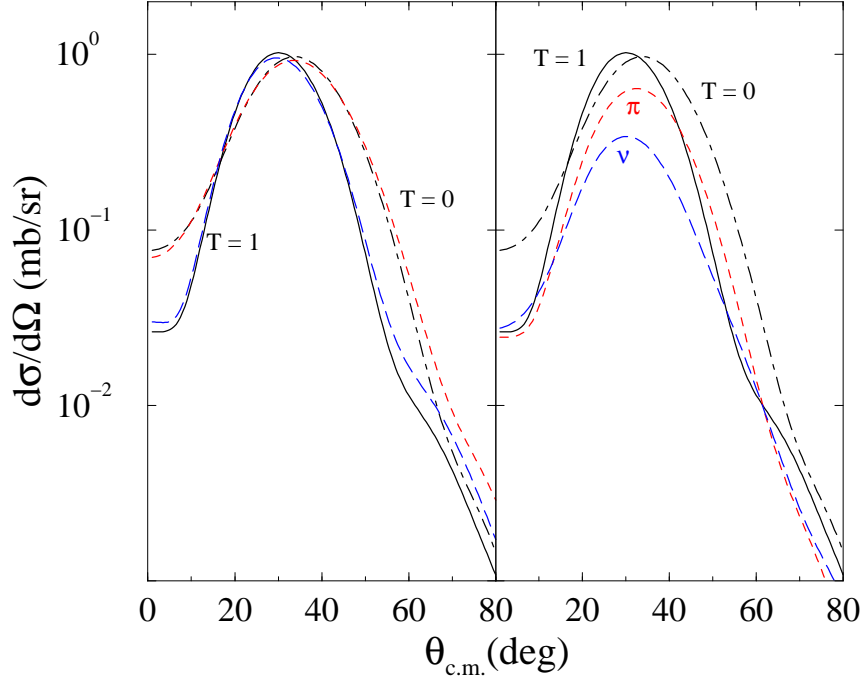}}
\caption{\label{Fig9-pure-pp-xs}(Color online)
Differential  cross  sections for the inelastic scattering of  135 
MeV  protons  from  ${}^{16}$O  exciting  pure isospin ``stretched'' 
$4^-$ states.} 
\end{figure}
%%%%%%%%%%%%%%%%%%%%%%%%%%%%%%%%%%%%%%
\begin{figure}
\scalebox{0.7}{\includegraphics*{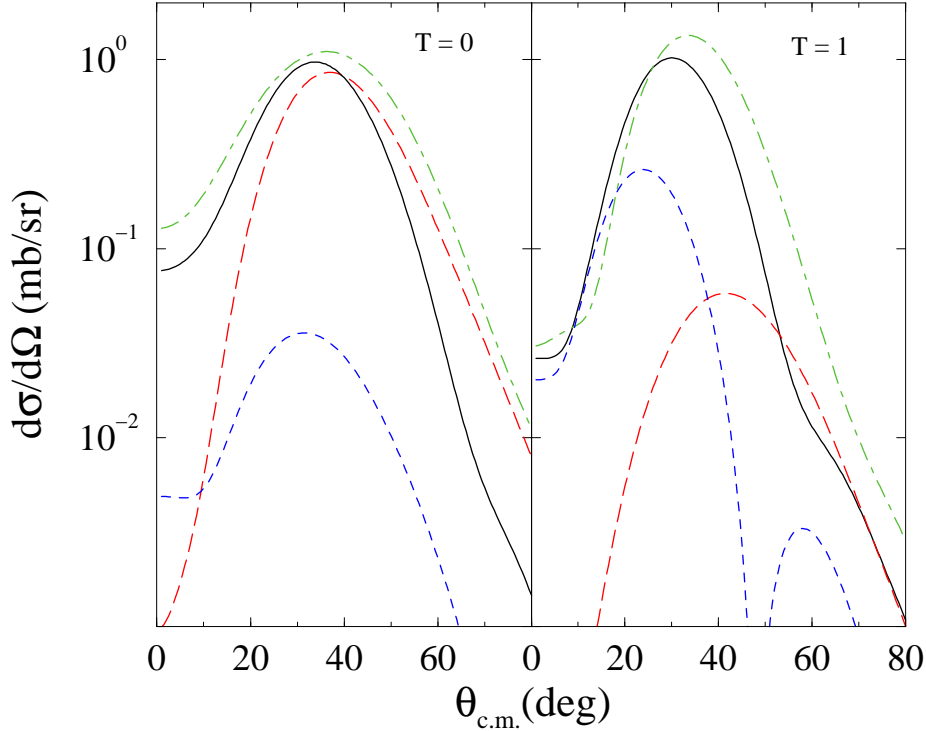}}
\caption{\label{Fig10-pp-comps}(Color online)
Differential cross sections for the inelastic scattering of 135 MeV 
protons from ${}^{16}$O  exciting pure isospin  ``stretched''  $4^-$ 
states   and   the  contributions from different components of the 
transition operator.} 
\end{figure}
%%%%%%%%%%%%%%%%%%%%%%%%%%%%%%%%%%%%%%

     First consider cross sections from the excitation of the pure 
isoscalar and isovector ``stretched'' $4^-$ states.   The results of 
those DWA calculations  are displayed in Fig~\ref{Fig9-pure-pp-xs}.
The isovector and isoscalar results are those labeled by $T=1$   and
$T=0$ in both panels of the figure.  Harmonic oscillator bound 
state functions with $b = 1.7$ fm were used to get most of those 
results.   Shown in the right hand panel are  the  pure  proton 
($\pi$) and pure neutron ($\nu$) excitation cross sections. 
They are noticeably different and the interferences between   them 
to effect the  pure isoscalar and pure isovector  excitations 
result in cross sections that have similar magnitude but different 
peak positions. This is a quite different phenomenon than is evident 
with the electron scattering form factors for which the  isoscalar
and isovector form factors  have so vastly different   magnitudes. 
In the left panel, the harmonic oscillator results    (solid curve for $T=1$, 
dot-dashed curve for $T=0$) are compared with those  obtained  using 
the WS bound state functions    (long dash curve for $T=1$,
short dash curve for $T=0$). Clearly the differences are very minor 
until larger momentum transfer values are involved. 

 The proton scattering results are determined by the nature of the 
transition  operator  and  the  characteristics  of  it  that most 
influence the results are revealed in Fig.~\ref{Fig10-pp-comps}.
   The cross sections from excitation of the pure  $4^-$ isoscalar 
(left) and isovector (right) states and the contributions from the 
central parts of the transition operator (dashed curves), from the 
two-nucleon spin-orbit (${\bf L\cdot S}$) components  (long dashed 
curves) and from the tensor (${\bf S}_{12}$) components (dot-dashed 
curves) are shown.   Clearly the dominant component in these $4^-$ 
excitations is that of the tensor force while the ${\bf L\cdot S}$ 
terms are more important for the isoscalar than for the  isovector 
transitions.               The interference between the tensor and 
${\bf L\cdot S}$ contributions in the isoscalar transition is  the 
cause  of  the  shape difference of the total result from that for 
the isovector excitation.

%%%%%%%%%%%%%%%%%%%%%%%%%%%%%%%%%%%%%%%%%%%%%%%%%%%%%%%%
\begin{figure}[ht]
\scalebox{0.6}{\includegraphics*{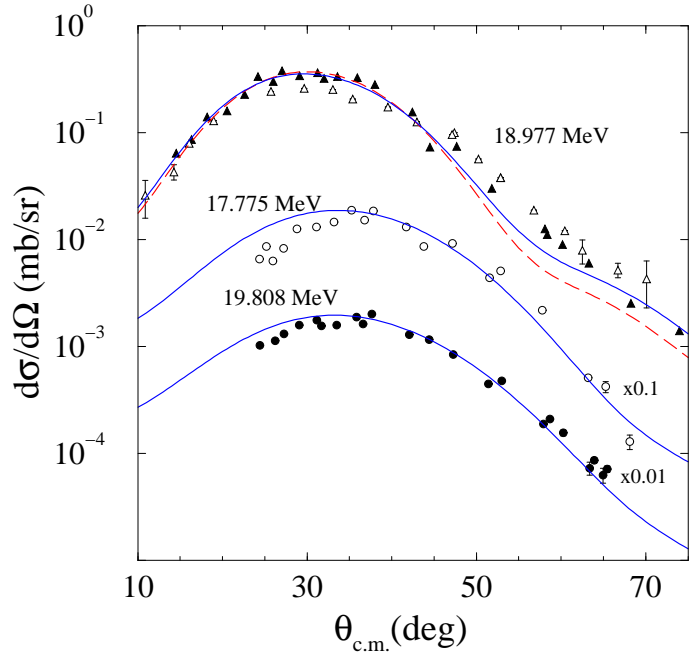}}
\caption{\label{Fig11-pp-135}(Color online)
Differential cross sections for the inelastic scattering of 135 MeV 
protons from ${}^{16}$O exciting the three $4^-$ states.
The 17.775 and 19.808~MeV data and results 
have been shifted by 0.1 and 0.01 in scale for visualization.   The
charge exchange (IAS) ($p,n$) cross section data divided by 2   are
shown by the open triangles.}
\end{figure}
%%%%%%%%%%%%%%%%%%%%%%%%%%%%%%%%%%%%%%%%%%%%%%%%%%%%%%%%
      The complete results for the 135 MeV proton scattering cross 
sections  to  the  actual  $4^-$ states in ${}^{16}$O are shown in 
comparison with data in Fig.~\ref{Fig11-pp-135}. The isospin mixing 
as suggested by the analyses  of the  pion and electron scattering 
data have been used and the results scaled to fit;    the scalings 
being identified as the configuration mixing weights. 
The solid lines are the results of DWA calculations made using the 
WS bound state functions while the dashed curve  displays 
the result for the 18.977 MeV excitation found using      harmonic 
oscillators.      Clearly the shapes of the measured data are well
reproduced with the   distinguishing    features of the dominantly 
isovector and isoscalar transitions being most evident.        The 
scales, $N_i$, of Eq.~(\ref{mixediso}) that were required to  obtain 
these matches to the cross-section magnitudes were   $N_1 = 0.42$, 
$N_2 = 0.62$, and $N_3 = 0.49$.

\begin{figure}
\scalebox{0.6}{\includegraphics*{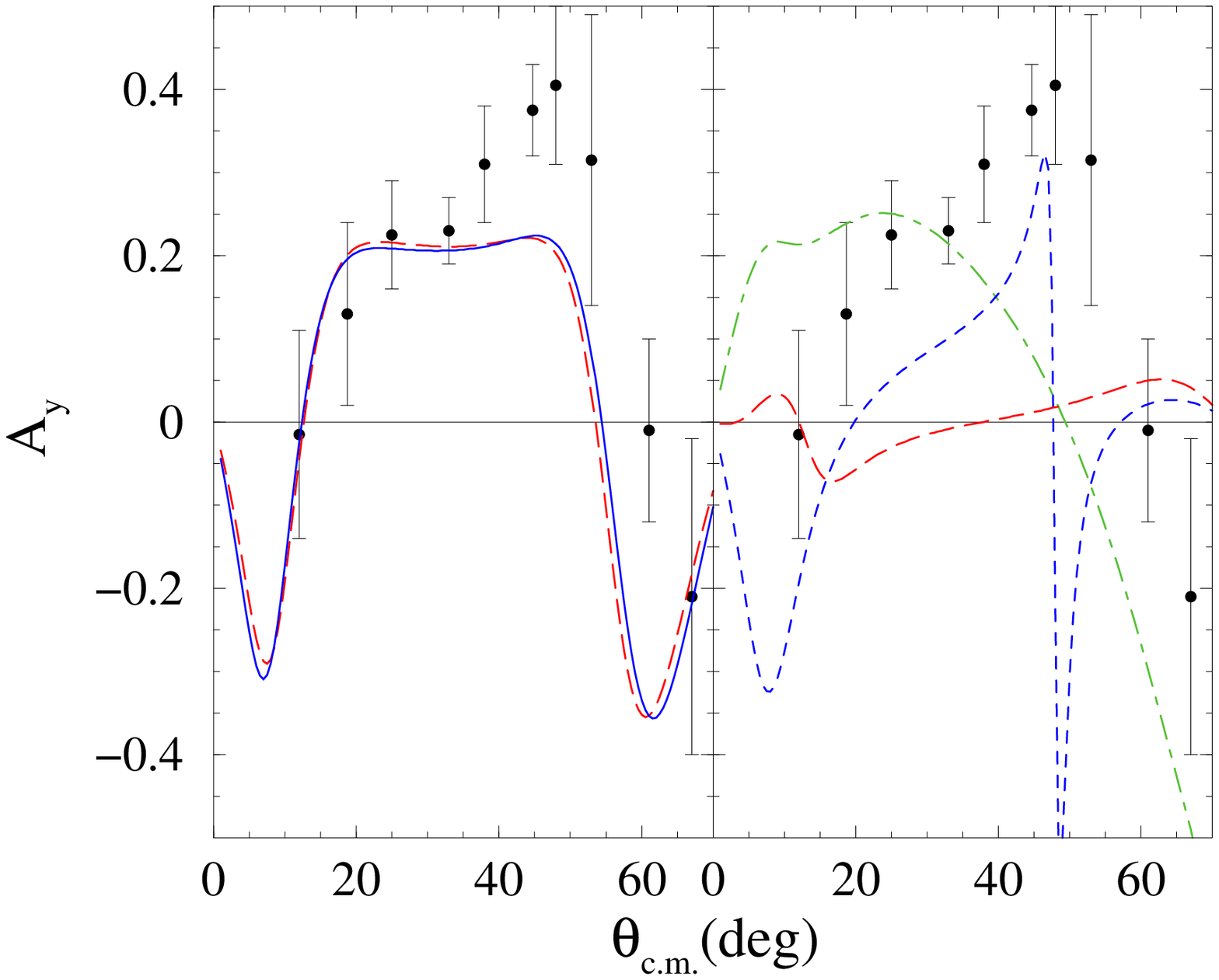}}
\caption{\label{Fig12-anal}(Color online)
Analyzing    power    for the charge exchange reaction of 135 MeV 
protons from ${}^{16}$O exciting the $4^-$ IAS state at 6.41~MeV.}
\end{figure}

  Finally we consider the analyzing power results to compare with 
the data taken in a 135 MeV charge exchange experiment~\cite{Ma82}.
The results are presented in Fig.~\ref{Fig12-anal}.
%%%%%%%%%%%%%%%%%%%%%%%%%%%%%%%%%%%%%%%%%%%%%%%%%%%%%%%%
%%%%%%%%%%%%%%%%%%%%%%%%%%%%%%%%%%%%%%
In the left panel the total result as found using the  WS 
bound state functions is displayed by the solid curve   while that 
found using the harmonic oscillators is depicted by  the    dashed 
curve.    In the right panel, the results found using the separate 
central (dash),  tensor  (dot-dash),  and  two-nucleon  spin-orbit 
(long dash) terms in the transition operator are presented.   Note 
that these component results are normalized against the individual 
component cross sections they form        (see the right panel of 
Fig.~\ref{Fig10-pp-comps}) and so their relative importance  should 
be considered with comparison of the component against the   total 
cross section.    However, the tensor and spin-orbit amplitudes in
particular are needed and their interference is crucial in finding
the final total result that matches the data as well as it does.

%%%%%%%%%%%%%%%%%%%%% CONCLUSIONS %%%%%%%%%%%%%%%%%%%%%%%%
\section{Conclusions}
\label{Conclu}

The simultaneous analyses of electron, pion, and proton scattering
data leading to the 4$^-$ states in ${}^{16}$O ascertain  that the
states at   17.775 MeV    and  19.808 MeV are dominantly isoscalar 
having isovector admixtures specified by
%%%%%%%%%%%%%%%%
\begin{equation}
N_1 \sin(\epsilon_1) = 0.42\ \sin(11.5^\circ)\ ,
\end{equation}
%%%%%%%%%%%%%%%%%%%%%%%%%%%%%%
and
%%%%%%%%%%%%%%%%%%%%%%
\begin{equation}
N_3 \sin(\epsilon_3) = 0.49\ \sin(-9.8^\circ)\ ,
\end{equation}
%%%%%%%%%%%%%%%%%%%%%%%
while  the  18.977 MeV state is virtually pure isovector in nature 
with $ N_2 = 0.62 $ being expected from the proton scattering data 
analyses.     Thus we assess that the fractional exhaustion of the 
($0d_{\frac{3}{2}}-0p_{\frac{3}{2}}^{-1}; T$) 
strengths exhausted in these transitions are
%%%%%%%%%%%%%%%%%%%%%%%%%%%%%%%%%%%%%%%%%%%%%%%%%%%%%%%%%%%%%%%%%%
\begin{eqnarray}
(0d_{\frac{3}{2}}-0p_{\frac{3}{2}}^{-1}; T=0)
&=& \left|0.42\ \cos(11.5^\circ)\right|^2  + 
\left|0.49\ \cos(-9.8^\circ)\right|^2 = 0.40
\nonumber\\
(0d_{\frac{3}{2}}-0p_{\frac{3}{2}}^{-1}; T=1)
&=& \left|0.62\right|^2 + \left|0.42\ \sin(11.5^\circ)\right|^2
 +\left| 0.49\ \sin(-9.8^\circ)\right|^2
= 0.40\ .
\end{eqnarray}
%%%%%%%%%%%%%%%%%%%%%%%%%%%%%%%%%%%%%%%%%%%%%%%%%%%%%%%%%%%%%%%%%%%
These values are less than one half of what has been assessed in the 
past from particle transfer data.  But as the transfer data assessment 
is very much phenomenological model calculation dependent, particularly
with what overall strength of interaction is chosen in zero-range DWBA
calculations, and with a limited model spectroscopy which ascribed 
$J^\pi = 2^-$ to the state at 17.775 MeV excitation.
As this is now known to be a $4^-$ state, and 
there is at least a fourth $4^-$ state at 20.5~MeV 
excitation, we believe our deduced numbers to be the more credible.

%%%%%%%%%%%%%%%%%%%%%%%%%%%%%%%%%%%%%%%%%%%%%%%%%%%%%%%%%%%%%%%%%%%%
\begin{acknowledgments}
This research was supported  by a research grant from the 
Australian Research Council and by a grant from the Cheju National 
University Development Foundation (2004).
\end{acknowledgments}
%%%%%%%%%%%%%%%%%%%%%%%%%%%%%%%%%%%%%%%%%%%%%%%%%%%%%%%%%%%%%%%%%%

%%%%%%%%%%%%%%%%%%%  APPENDICES %%%%%%%%%%%%%%%%%%%%%%%%%%%%%
\appendix
\section{Normalization of the $\left| 4^-; T_f \right>$ states}
\label{App-norm}

The normalization constant ${\cal N}$ of the $4^-; T_f$ states follows from 
%%%%%%%%%%%%%%%%%%%%%
\begin{eqnarray}
1 &=& \left< 4^-; T_f \Bigl| 4^-; T_f \right>\nonumber\\
&=& {\cal N}^2 \sum 
(-)^{\left(\frac{1}{2} - \alpha \right)}
(-)^{\left(\frac{1}{2} - \alpha^\prime \right)}
(-)^{\left(\frac{3}{2} - m_3 \right)}
(-)^{\left(\frac{3}{2} - m_3^\prime \right)}
\nonumber\\
&&\times
\left\langle \frac{1}{2}\, \frac{1}{2}\, \alpha\, -\alpha \Bigl| T\, 0 
\right\rangle
\left\langle \frac{1}{2}\, \frac{1}{2}\, \alpha^\prime\, -\alpha^\prime \Bigl| 
T\, 0 \right\rangle
\left\langle \frac{3}{2}\, \frac{5}{2}\, m_3\, -m_5 \Bigl| 4\, -M_4 
\right\rangle
\left\langle \frac{3}{2}\, \frac{5}{2}\, m_3^\prime\, -m_5^\prime \Bigl| 4\, 
-M_4 \right\rangle
\nonumber\\
&& \times
\left\langle 0^+ \Bigl| 
a_{\frac{3}{2} m_3^\prime \alpha^\prime}^\dagger 
a_{\frac{5}{2} m_5^\prime \alpha^\prime} 
a_{\frac{5}{2} m_5 \alpha}^\dagger 
a_{\frac{3}{2} m_3 \alpha} \Bigr| 0^+\right\rangle\ .
\end{eqnarray}
%%%%%%%%%%%%%%%%%%%%
The ground state expectation values then reduce as
%%%%%%%%%%%%%%%%%%%
\begin{eqnarray}
\left\langle 0^+ \Bigl|
a_{\frac{3}{2} m_3^\prime \alpha^\prime}^\dagger
a_{\frac{5}{2} m_5^\prime \alpha^\prime}
a_{\frac{5}{2} m_5 \alpha}^\dagger
a_{\frac{3}{2} m_3 \alpha} \Bigr| 0^+\right\rangle 
&=& \left\langle 0^+ \Bigl| \left\{ \delta_{m_5 m_5^\prime} 
- a_{\frac{5}{2} m_5 \alpha}^\dagger
a_{\frac{5}{2} m_5^\prime \alpha^\prime}\right\}
a_{\frac{3}{2} m_3^\prime \alpha^\prime}^\dagger
a_{\frac{3}{2} m_3 \alpha} \Bigr| 0^+\right\rangle 
\nonumber\\
&=&
\delta_{m_5^\prime m_5} \delta_{m_3^\prime m_3} 
\delta_{\alpha^\prime \alpha} \left\{1 - \sigma_{\frac{5}{2} \alpha}\right\} 
\sigma_{\frac{3}{2} \alpha}\ ,
\end{eqnarray}
%%%%%%%%%%%%%%%%%%%
where the fractional occupancies (of each orbit) are defined from
%%%%%%%%%%%%%%%%%%%%
\begin{equation}
\left\langle 0^+ \Bigl| a_{j m^\prime \alpha^\prime}^\dagger
a_{j m \alpha} \Bigr| 0^+ \right\rangle
= \delta_{ m^\prime m} \delta_{\alpha^\prime \alpha}\ 
\sigma_{j \alpha}\ .
\label{frac-occ}
\end{equation}
%%%%%%%%%%%%%%%%%%%
In this we assume that every orbit $j, m$ is equally occupied with
%%%%%%%%%%%%%%%%%%%%%%%%
\begin{equation}
\sigma_{j x} = \sigma_{j m x} = \frac{1}{2j + 1} n_{jx}\ ,
\end{equation}
%%%%%%%%%%%%%%%%%%%%%%%
where $n_{j x}$ is the number of nucleons of type $x$ that are contained
in the shell $j$ in the target.   For a packed shell, as $n_{j x} =
(2j + 1)$, then $\sigma_{j x} = 1$.

Thus 
%%%%%%%%%%%%%%%%%%%%%%%
\begin{eqnarray}
{\cal N}^{-2} &=&
\sum_{m_3 m_5} \left[
\left\langle \frac{3}{2}\, \frac{5}{2}\, m_3\, -m_5 \Bigl| 4\, -M_4 
\right\rangle
\right]^2
\sum_{\alpha}
\left[\left\langle \frac{1}{2}\, \frac{1}{2}\, \alpha\, -\alpha \Bigl| T\, 
0 \right\rangle \right]^2
\left\{1 - \sigma_{\frac{5}{2} \alpha} \right\} \sigma_{\frac{3}{2} \alpha}
\nonumber\\
&=&
\sum_{\alpha}
\left[\left\langle \frac{1}{2}\, \frac{1}{2}\, \alpha\, -\alpha \Bigl| T\, 
0 \right\rangle \right]^2
\left\{1 - \sigma_{\frac{5}{2} \alpha} \right\} \sigma_{\frac{3}{2} \alpha}
\; =\; \frac{1}{2} 
\sum_{\alpha} \left\{1 - \sigma_{\frac{5}{2} \alpha} \right\} 
\sigma_{\frac{3}{2} \alpha}\ ,
\end{eqnarray}
%%%%%%%%%%%%%%%%%%%%%%%%
and then, with charge symmetry assumed, the sum over bound nucleon isospins
$\alpha$ gives just a factor of 2 and, 
%%%%%%%%%%%%%%%%%%%%%%%
\begin{equation}
{\cal N} = {\sqrt 2} 
\left[\sum_{\alpha} \left\{1 - \sigma_{\frac{5}{2} \alpha} \right\} 
\sigma_{\frac{3}{2} \alpha} \right]^{-\frac{1}{2}}\ 
\equiv \frac{1}
{\sqrt{\left\{1 - \sigma_{\frac{5}{2}} \right\} \sigma_{\frac{3}{2}}} }
\label{Normeq}
\end{equation}
%%%%%%%%%%%%%%%%%%%%%%%

\section{Spectroscopic amplitudes}

We consider the general matrix element, from which the spectroscopic
amplitudes to be used in DWBA98 code evaluations can be derived, namely
%%%%%%%%%%%%%%%
\begin{equation}
{\cal M}_{(x,y)} = 
\left\langle 4^-; T_f \left|
\left[a_{j_2 x}^\dagger \otimes {\tilde a}_{j_1 y} \right]^{(I)}_{(M_4)}
\right| 0^+; 0\right\rangle\ ,
\label{gen-excit}
\end{equation}
%%%%%%%%%%%%%%%%
where now we have
%%%%%%%%%%%%%%%%
\begin{equation}
\left[a_{j_2 x}^\dagger \otimes {\tilde a}_{j_1 y} \right]^{(I)}_{(N)}
= \sum_{m_1 m_2} (-)^{\left(j_1 - m_1\right)}
\left\langle j_1\, j_2\, m_1\, -m_2 \Bigl| I\, -N\right\rangle \ 
a_{j_2 m_2 x}^\dagger\, a_{j_1 m_1 y}
\end{equation}
%%%%%%%%%%%%%%%
while
%%%%%%%%%%%%%%%%%%%
\begin{equation}
\left|4^-;T_f\right>= {\cal N} \left[ 
a^\dagger_{\left(\frac{5}{2},\frac{1}{2}\right)}
\times {\tilde a}^{\phantom{\dagger}}_{\left(\frac{3}{2},\frac{1}{2}\right)}  
\right]^{(4,T_f)}_{(M_4, 0)} \left|0^+;0\right>\ ,
\end{equation}
%%%%%%%%%%%%%%%%%
with normalization as given in Eq.~(\ref{Normeq}).

Notice that the transition operator is coupled only in angular momentum 
and not in isospin.  That is so we find spectroscopic amplitudes for 
bound proton and bound neutron excitations separately as are required 
in evaluations of both electron form factors and cross 
sections initiated by nucleons. Those amplitudes, in fact, are the 
reduced matrix elements from Eq.~(\ref{gen-excit}),  
%%%%%%%%%%%%%%%%%
\begin{equation}
{\cal M}_{(x,y)} = \frac{1}{3} \left< 0\, I\, 0\, N \Bigl| 4\, M_f \right>\  
\left<4^-; T_f \left\| 
\left[a_{j_2 x}^\dagger \otimes {\tilde a}_{j_1 y} \right]^I
 \right\| 0^+; 0\right> \:
=\; \frac{1}{3}\, \delta_{I 4}\, \delta_{N M_f}\, S_{j_1\, j_2\, 4}^{(x,y)}\ .
\end{equation}
%%%%%%%%%%%%%%%%% 
Thus, evaluation of the spectroscopic amplitudes for these transitions
is made using the inverse of this, namely
$S_{j_1 j_2 4}^{(x,y)} = 3  \delta_{I 4} \delta_{N M_f} {\cal M}_{(x,y)}$.

For inelastic scattering, the nucleon type does not change
so the $x = y$ in the above.  However to consider charge exchange 
reactions, for which $x = -y$, we have to revise the specification
of the final nuclear state. That transition is to the IAS
which relates to the $T=1$ state in ${}^{16}$O by the action of an extra
isospin projection changing operator.

\subsection{Inelastic scattering}
\label{inel-Samps}

With $x = y$, the general matrix element expands as
%%%%%%%%%%%%%%%%%%
\begin{eqnarray}
{\cal M}_{(x)} &=&
\left[ \left\{1 - \sigma_{\frac{5}{2}} 
\right\} \sigma_{\frac{3}{2}}\right]^{-\frac{1}{2}}
\sum_{m_3 m_5 \alpha m_1 m_2} 
(-)^{\left( \frac{1}{2} - \alpha\right)}
\left\langle \frac{1}{2}\, \frac{1}{2}\, \alpha\, -\alpha \Bigl| T_f\, 0
\right\rangle
\nonumber\\
&&\;\; \times
(-)^{\left( \frac{3}{2} - m_3\right)}
\left\langle \frac{3}{2}\, \frac{5}{2}\, m_3\, -m_5 \Bigl| 4\, -M_4
\right\rangle
\, (-)^{\left(j_1 - m_1\right)} 
\left\langle j_1\, j_2\, m_1\, -m_2 \Bigl| I\, -N\right\rangle 
\nonumber\\
&&\;\; \times
\left\langle 0^+; 0 \left|
a^\dagger_{\frac{5}{2} m_5 \alpha}
a_{\frac{3}{2} m_3 \alpha}
a_{j_2 m_2 x}^\dagger a_{j_1 m_1 x}
\right| 0^+; 0\right\rangle\ .
\end{eqnarray}
%%%%%%%%%%%%%%%%%%%
Then using the fractional occupancy representation of the expectation, 
Eq.~(\ref{frac-occ}), this reduces to
%%%%%%%%%%%%%%%%%%%
\begin{eqnarray}
{\cal M}_{(x)} &=&
\sqrt{ \left\{1 - \sigma_{\frac{5}{2}} \right\} \sigma_{\frac{3}{2}} }
\ (-)^{\left(\frac{1}{2} - x\right)}
\left<\frac{1}{2}\, \frac{1}{2}\, x\, -x \Bigl| T_f\, 0 \right>
\nonumber\\
&& \times 
\sum_{m_1 m_2 m_3 m_5} \delta_{j_1 \frac{3}{2}} \delta_{m_1 m_3}
\delta_{j_2 \frac{5}{2}} \delta_{m_2 m_5}
\left< j_1\, j_2\, m_1\, -m_2 \Bigl| I\, -N \right>
\left< \frac{3}{2}\, \frac{5}{2}\, m_3\, -m_5 \Bigl| 4\, -M_f \right>
\nonumber\\
&=& \sqrt{ \left\{1 - \sigma_{\frac{5}{2}} \right\} \sigma_{\frac{3}{2}} }
\ (-)^{\left(\frac{1}{2} - x\right)}
\left<\frac{1}{2}\, \frac{1}{2}\, x\, -x \Bigl| T_f\, 0 \right>
\nonumber\\
&=&
 \sqrt{ \left\{1 - \sigma_{\frac{5}{2}} \right\} \sigma_{\frac{3}{2}} }\; 
\frac{1}{\sqrt 2}\left[\delta_{T_f 0} + (-)^{\left(\frac{1}{2} - x\right)}
\delta_{T_f 1} \right]\ ,
\end{eqnarray}
%%%%%%%%%%%%%%%%%%
so that, for inelastic scattering,
%%%%%%%%%%%%%%%
\begin{equation}
S_{j_1 j_2 4}^{(x)} = \delta_{j_1 \frac{3}{2}} \delta_{j_2 \frac{5}{2}}
\frac{3}{\sqrt 2}\  
\sqrt{ \left\{1 - \sigma_{\frac{5}{2}} \right\} \sigma_{\frac{3}{2}} }  
\left[\delta_{T_f 0} + (-)^{\left(\frac{1}{2} - x\right)}
\delta_{T_f 1} \right]\ .
\label{Sinel}
\end{equation}
%%%%%%%%%%%%%%%

\subsection{Charge exchange to the IAS $4^-$ in ${}^{16}$F}
\label{pn-Samps}

First we need to specify the form that the IAS takes.
With $T^- = \sum_{jm} a_{jm-\frac{1}{2}}^\dagger a_{jm \frac{1}{2}}$ 
being the operator that changes a neutron to a proton, the $4^-$
state in ${}^{16}$F that is an IAS to the particle-hole model isovector
$4^-$ state in ${}^{16}$O has the form
%%%%%%%%%%%%%%%%%%%%%
\begin{eqnarray}
\left| 4^-; IAS\right\rangle &=&
\left| 4^-; T_f = 1, M_{T_f} = -1 \right\rangle
= {\cal P} T^- 
\left[ 
a^\dagger_{\left(\frac{5}{2} \frac{1}{2}\right)} \otimes 
{\tilde a}_{\left(\frac{3}{2} \frac{1}{2}\right)}
\right]^{(4,T_f =1)}_{(M_4, 0)}
\Bigr| 0^+; 0 \Bigr>
\nonumber\\
&=& {\cal P} \sum_{m, \alpha, m_3 m_5} 
(-)^{\left(\frac{1}{2} - \alpha \right)} 
\left< \frac{1}{2}\, \frac{1}{2}\, \alpha\, -\alpha \Bigl| 
T_f(=1)\, 0 \right>
(-)^{\left(\frac{1}{2} - m_3 \right)}
\left<\frac{3}{2}\, \frac{5}{2}\, m_3\, -m_5 \Bigl| 4\, -M_f \right>
\nonumber\\
&&\;\;\; \times
\ a^\dagger_{jm-\frac{1}{2}} a_{jm \frac{1}{2}} 
a^\dagger_{\frac{5}{2} m_5 \alpha} a_{\frac{3}{2} m_3 \alpha} \Bigl|0^+;0
\Bigr>\ ,
\end{eqnarray}
%%%%%%%%%%%%%%%%%%%%%
where $\cal P$ is the normalization.
Then, the particle-hole description of the $4^-$ states requires
%%%%%%%%%%%%%%%%%%%
\begin{equation}
a^\dagger_{jm-\frac{1}{2}} a_{jm \frac{1}{2}}
a^\dagger_{\frac{5}{2} m_5 \alpha} a_{\frac{3}{2} m_3 \alpha} \Bigl|0^+;0
\Bigr> \equiv
\delta_{\alpha, \frac{1}{2}} \delta_{j, \frac{5}{2}} \delta_{m m_5}
a^\dagger_{\frac{5}{2} m_5 -\frac{1}{2}} a_{\frac{3}{2} m_3 \frac{1}{2}}
\left[
1 - a^\dagger_{\frac{5}{2}, m_5 \frac{1}{2}} a_{\frac{5}{2} m_5 \frac{1}{2}} 
\right] 
\Bigl| 0^+;0 \Bigr>\ ,
\end{equation}
%%%%%%%%%%%%%%%%
and the isospin terms are simply
%%%%%%%%%%%%%%%%%%
\begin{equation}
\delta_{\alpha \frac{1}{2}}\ 
(-)^{\left(\frac{1}{2} - \alpha \right)}
\left< \frac{1}{2}\, \frac{1}{2}\, \alpha\, -\alpha \Bigl| T_f(=1)\; 0 \right>
= \frac{1}{\sqrt 2}\ ,
\end{equation}
%%%%%%%%%%%%%%%%%%
the IAS state becomes
%%%%%%%%%%%%%%%%%
\begin{eqnarray}
\Bigl| 4^-; IAS\Bigr> &=&
{\cal P} \frac{1}{\sqrt 2} \sum_{m_3 m_5} 
(-)^{\left(\frac{1}{2} - m_3 \right)}
\left<\frac{3}{2}\, \frac{5}{2}\, m_3\, -m_5 \Bigl| 4\, -M_f \right>
\nonumber\\
&&\;\;\; \times
a^\dagger_{\frac{5}{2} m_5 -\frac{1}{2}} a_{\frac{3}{2} m_3 \frac{1}{2}}
\left[
1 - a^\dagger_{\frac{5}{2}, m_5 \frac{1}{2}} a_{\frac{5}{2} m_5 \frac{1}{2}}
\right] 
\Bigl| 0^+;0 \Bigr>\ .
\label{IASeq}
\end{eqnarray}
%%%%%%%%%%%%%%%%%

Normalization then defines $\cal P$ since, with $a_{jmx}^\dagger a_{jmx}$ is 
the '$x$' nucleon number operator leading to $\sigma_{jx}$
%%%%%%%%%%%%%%%%%% 
\begin{eqnarray}
1 &=&\left< 4^-; IAS\bigl|4^-; IAS \right>
\nonumber\\
&=& {\cal P}^2 \frac{1}{2} 
\sum_{m_3 m_5 m_3^\prime m_5^\prime}
(-)^{\left(\frac{1}{2} - m_3^\prime \right)}
\left<\frac{3}{2}\, \frac{5}{2}\, m_3^\prime\, -m_5^\prime \Bigl| 
4\, -M_f \right>
(-)^{\left(\frac{1}{2} - m_3 \right)}
\left<\frac{3}{2}\, \frac{5}{2}\, m_3\, -m_5 \Bigl| 4\, -M_f \right>
\nonumber\\
&&\times
\left< 0^+; 0\left|
\left[1 - \sigma_{\frac{5}{2}\frac{1}{2}} \right]
a^\dagger_{\frac{3}{2} m_3^\prime \frac{1}{2}}
a_{\frac{5}{2} m_5^\prime -\frac{1}{2}}
a^\dagger_{\frac{5}{2} m_5 -\frac{1}{2}} a_{\frac{3}{2} m_3 \frac{1}{2}}
\left[1 - \sigma_{\frac{5}{2}\frac{1}{2}} \right]
\right| 0^+; 0 \right>
\nonumber\\
&=&
{\cal P}^2 \frac{1}{2} \left[1 - \sigma_{\frac{5}{2} \frac{1}{2}}\right]^2
\left[1 - \sigma_{\frac{5}{2} -\frac{1}{2}} \right] \ 
\sigma_{\frac{3}{2} \frac{1}{2}}
\end{eqnarray}
%%%%%%%%%%%%%%%%%%%
so that on inversion and with identical fractional occupancies for 
proton and neutron shells,
%%%%%%%%%%%%%%%%%
\begin{equation}
{\cal P} = 
\sqrt{ 2\Bigl/ 
\left\{ \left[1 - \sigma_{\frac{5}{2}} \right]^3 \sigma_{\frac{3}{2}}
\right\}} .
\label{IASnorm}
\end{equation}
%%%%%%%%%%%%%%%%%%%

The spectroscopic amplitude follows then by using 
Eqs.~(\ref{IASeq}) and (\ref{IASnorm}) in forming
%%%%%%%%%%%%%%%%%%%%
\begin{eqnarray}
S_{j_1 j_2 I}^{-\frac{1}{2} \frac{1}{2}} &=&
3 \left< 4^-; IAS \left| 
\left[ a_{j_2 -\frac{1}{2}} \otimes {\tilde a}_{j_1 \frac{1}{2}} 
\right]^{I}_{N}
\right|0;0\right>
\nonumber\\
&=& \frac{3}{\sqrt 2} 
\left[
\frac{2}{\left[1 - \sigma_{\frac{5}{2}} \right]^3 \sigma_{\frac{3}{2}} }
\right]^{\frac{1}{2}}
\nonumber\\
&& \times \sum_{m_1 m_2}
(-)^{(j_1 - m_1)} \left< j_1\, j_2\, m_1\, -m_2 \Bigl|I\, -N \right>
\,(-)^{(\frac{3}{2} - m_3)} 
\left<\frac{3}{2}\, \frac{5}{2}\, m_3\, -m_5| 4\, -M_f\right>
\nonumber\\
&&\;\;\; \times
\left< 0^+; 0\left|
\left[1 - \sigma_{\frac{5}{2}}\right] 
a^\dagger_{\frac{3}{2} m_3 \frac{1}{2}}\ a_{\frac{5}{2} m_5 -\frac{1}{2}}
a^\dagger_{j_2 m_2 -\frac{1}{2}} a_{j_1 m_1 \frac{1}{2}}
\right| 0^+; 0\right>
\nonumber\\
&=& 3\frac{1}{\sqrt{\left[1 - \sigma_{\frac{5}{2}} 
\right]^3 \sigma_{\frac{3}{2}}}}
\left[ 1 - \sigma_{\frac{5}{2}}\right]^2 \sigma_{\frac{3}{2}}
\; =\;  
3 \sqrt{\left[ 1 - \sigma_{\frac{5}{2}}\right] \sigma_{\frac{3}{2}}} 
\; (\delta_{T_f 1})\ .
\end{eqnarray}
%%%%%%%%%%%%%%%%%%%%%%%%%%%%%%%%%%
This is $\sqrt 2$ times the value found for inelastic scattering as given 
in Eq.~(\ref{Sinel}) and so the charge exchange cross section should be 
twice that for inelastic scattering to the IAS.

%%%%%%%%%%%%%%%%%%%%%%%%%%%%%%%%%%%%%%%%%%%%%%%%%%%%%%%%%%%%%%%%%
\bibliography{oxy}

\begin{thebibliography}{30}
\expandafter\ifx\csname natexlab\endcsname\relax\def\natexlab#1{#1}\fi
\expandafter\ifx\csname bibnamefont\endcsname\relax
  \def\bibnamefont#1{#1}\fi
\expandafter\ifx\csname bibfnamefont\endcsname\relax
  \def\bibfnamefont#1{#1}\fi
\expandafter\ifx\csname citenamefont\endcsname\relax
  \def\citenamefont#1{#1}\fi
\expandafter\ifx\csname url\endcsname\relax
  \def\url#1{\texttt{#1}}\fi
\expandafter\ifx\csname urlprefix\endcsname\relax\def\urlprefix{URL }\fi
\providecommand{\bibinfo}[2]{#2}
\providecommand{\eprint}[2][]{\url{#2}}

\bibitem[{\citenamefont{Uberall}(1971)}]{Ub71}
\bibinfo{author}{\bibfnamefont{H.}~\bibnamefont{Uberall}},
  \emph{\bibinfo{title}{Electron Scattering from Complex Nuclei}}
  (\bibinfo{publisher}{Academic Press}, \bibinfo{address}{New York},
  \bibinfo{year}{1971}).

\bibitem[{\citenamefont{Friar and Fallieros}(1984)}]{Fr84}
\bibinfo{author}{\bibfnamefont{J.~L.} \bibnamefont{Friar}} \bibnamefont{and}
  \bibinfo{author}{\bibfnamefont{S.}~\bibnamefont{Fallieros}},
  \bibinfo{journal}{Phys. Rev. C} \textbf{\bibinfo{volume}{29}},
  \bibinfo{pages}{1645} (\bibinfo{year}{1984}).

\bibitem[{\citenamefont{Friar and Fallieros}(1985)}]{Fr85}
\bibinfo{author}{\bibfnamefont{J.~L.} \bibnamefont{Friar}} \bibnamefont{and}
  \bibinfo{author}{\bibfnamefont{S.}~\bibnamefont{Fallieros}},
  \bibinfo{journal}{Phys. Rev. C} \textbf{\bibinfo{volume}{31}},
  \bibinfo{pages}{2027} (\bibinfo{year}{1985}).

\bibitem[{\citenamefont{Karataglidis
  et~al.}(1995{\natexlab{a}})\citenamefont{Karataglidis, Halse, and
  Amos}}]{Ka95}
\bibinfo{author}{\bibfnamefont{S.}~\bibnamefont{Karataglidis}},
  \bibinfo{author}{\bibfnamefont{P.}~\bibnamefont{Halse}}, \bibnamefont{and}
  \bibinfo{author}{\bibfnamefont{K.}~\bibnamefont{Amos}},
  \bibinfo{journal}{Phys. Rev. C} \textbf{\bibinfo{volume}{51}},
  \bibinfo{pages}{2494} (\bibinfo{year}{1995}{\natexlab{a}}).

\bibitem[{\citenamefont{Amos et~al.}(2000)\citenamefont{Amos, Dortmans, von
  Geramb, Karataglidis, and Raynal}}]{Am00}
\bibinfo{author}{\bibfnamefont{K.}~\bibnamefont{Amos}},
  \bibinfo{author}{\bibfnamefont{P.~J.} \bibnamefont{Dortmans}},
  \bibinfo{author}{\bibfnamefont{H.~V.} \bibnamefont{von Geramb}},
  \bibinfo{author}{\bibfnamefont{S.}~\bibnamefont{Karataglidis}},
  \bibnamefont{and} \bibinfo{author}{\bibfnamefont{J.}~\bibnamefont{Raynal}},
  \bibinfo{journal}{Adv. in Nucl. Phys.} \textbf{\bibinfo{volume}{25}},
  \bibinfo{pages}{275} (\bibinfo{year}{2000}).

\bibitem[{\citenamefont{Hyde-Wright et~al.}(1987)}]{Hy87}
\bibinfo{author}{\bibfnamefont{C.~E.} \bibnamefont{Hyde-Wright}}
  \bibnamefont{et~al.}, \bibinfo{journal}{Phys. Rev. C}
  \textbf{\bibinfo{volume}{35}}, \bibinfo{pages}{880} (\bibinfo{year}{1987}).

\bibitem[{\citenamefont{Clausen et~al.}(1988)\citenamefont{Clausen, Peterson,
  and Lindgren}}]{Cl88}
\bibinfo{author}{\bibfnamefont{B.~L.} \bibnamefont{Clausen}},
  \bibinfo{author}{\bibfnamefont{R.~J.} \bibnamefont{Peterson}},
  \bibnamefont{and} \bibinfo{author}{\bibfnamefont{R.~A.}
  \bibnamefont{Lindgren}}, \bibinfo{journal}{Phys. Rev. C}
  \textbf{\bibinfo{volume}{38}}, \bibinfo{pages}{589} (\bibinfo{year}{1988}).

\bibitem[{\citenamefont{Lagoyannis et~al.}(2001)}]{La01}
\bibinfo{author}{\bibfnamefont{A.}~\bibnamefont{Lagoyannis}}
  \bibnamefont{et~al.}, \bibinfo{journal}{Phys. Lett.}
  \textbf{\bibinfo{volume}{B518}}, \bibinfo{pages}{27} (\bibinfo{year}{2001}).

\bibitem[{\citenamefont{Karataglidis et~al.}(2002)\citenamefont{Karataglidis,
  Amos, Brown, and Deb}}]{Ka02}
\bibinfo{author}{\bibfnamefont{S.}~\bibnamefont{Karataglidis}},
  \bibinfo{author}{\bibfnamefont{K.}~\bibnamefont{Amos}},
  \bibinfo{author}{\bibfnamefont{B.~A.} \bibnamefont{Brown}}, \bibnamefont{and}
  \bibinfo{author}{\bibfnamefont{P.~K.} \bibnamefont{Deb}},
  \bibinfo{journal}{Phys. Rev. C} \textbf{\bibinfo{volume}{65}},
  \bibinfo{pages}{044306} (\bibinfo{year}{2002}).

\bibitem[{\citenamefont{Dupuis et~al.}(2005)\citenamefont{Dupuis, Karataglidis,
  Bauge, Delaroche, and Gogny}}]{Ka05}
\bibinfo{author}{\bibfnamefont{M.}~\bibnamefont{Dupuis}},
  \bibinfo{author}{\bibfnamefont{S.}~\bibnamefont{Karataglidis}},
  \bibinfo{author}{\bibfnamefont{E.}~\bibnamefont{Bauge}},
  \bibinfo{author}{\bibfnamefont{J.~P.} \bibnamefont{Delaroche}},
  \bibnamefont{and} \bibinfo{author}{\bibfnamefont{D.}~\bibnamefont{Gogny}}
  (\bibinfo{year}{2005}), \bibinfo{note}{nucl-th/0504013}.

\bibitem[{\citenamefont{Pearlman}(1993)}]{Pe93}
\bibinfo{author}{\bibfnamefont{S.}~\bibnamefont{Pearlman}},
  \emph{\bibinfo{title}{ENDF/HE-VI Mat-625}} (\bibinfo{year}{1993}),
  \bibinfo{note}{BNL-48035}.

\bibitem[{\citenamefont{Anderson et~al.}(1979)}]{An79}
\bibinfo{author}{\bibfnamefont{B.~D.} \bibnamefont{Anderson}}
  \bibnamefont{et~al.}, \emph{\bibinfo{title}{IUCF annual report}}
  (\bibinfo{year}{1979}), \bibinfo{note}{p. 29}.

\bibitem[{\citenamefont{Madey et~al.}(1982)}]{Ma82}
\bibinfo{author}{\bibfnamefont{R.}~\bibnamefont{Madey}} \bibnamefont{et~al.},
  \bibinfo{journal}{Phys. Rev. C} \textbf{\bibinfo{volume}{25}},
  \bibinfo{pages}{1715} (\bibinfo{year}{1982}).

\bibitem[{\citenamefont{Ohnuma et~al.}(1982)}]{Oh82}
\bibinfo{author}{\bibfnamefont{H.}~\bibnamefont{Ohnuma}} \bibnamefont{et~al.},
  \bibinfo{journal}{Phys. Lett.} \textbf{\bibinfo{volume}{112B}},
  \bibinfo{pages}{206} (\bibinfo{year}{1982}).

\bibitem[{\citenamefont{Holtkamp et~al.}(1980)}]{Ho80}
\bibinfo{author}{\bibfnamefont{D.~B.} \bibnamefont{Holtkamp}}
  \bibnamefont{et~al.}, \bibinfo{journal}{Phys. Rev. Letts.}
  \textbf{\bibinfo{volume}{45}}, \bibinfo{pages}{420} (\bibinfo{year}{1980}).

\bibitem[{\citenamefont{Millener and Kurath}(1975)}]{Mi75}
\bibinfo{author}{\bibfnamefont{D.~J.} \bibnamefont{Millener}} \bibnamefont{and}
  \bibinfo{author}{\bibfnamefont{D.}~\bibnamefont{Kurath}},
  \bibinfo{journal}{Nucl. Phys.} \textbf{\bibinfo{volume}{A255}},
  \bibinfo{pages}{315} (\bibinfo{year}{1975}), \bibinfo{note}{and references
  cited therein}.

\bibitem[{\citenamefont{Smith et~al.}(1978)\citenamefont{Smith, Morton,
  Morrison, and Amos}}]{Sm78}
\bibinfo{author}{\bibfnamefont{R.}~\bibnamefont{Smith}},
  \bibinfo{author}{\bibfnamefont{J.}~\bibnamefont{Morton}},
  \bibinfo{author}{\bibfnamefont{I.}~\bibnamefont{Morrison}}, \bibnamefont{and}
  \bibinfo{author}{\bibfnamefont{K.}~\bibnamefont{Amos}},
  \bibinfo{journal}{Aust. J. Phys.} \textbf{\bibinfo{volume}{31}},
  \bibinfo{pages}{1} (\bibinfo{year}{1978}).

\bibitem[{\citenamefont{Mairle et~al.}(1978)\citenamefont{Mairle, Wagner, Doll,
  T.Knoepfle, and Breuer}}]{Ma78}
\bibinfo{author}{\bibfnamefont{G.}~\bibnamefont{Mairle}},
  \bibinfo{author}{\bibfnamefont{G.~J.} \bibnamefont{Wagner}},
  \bibinfo{author}{\bibfnamefont{P.}~\bibnamefont{Doll}},
  \bibinfo{author}{\bibfnamefont{K.}~\bibnamefont{T.Knoepfle}},
  \bibnamefont{and} \bibinfo{author}{\bibfnamefont{H.}~\bibnamefont{Breuer}},
  \bibinfo{journal}{Nucl. Phys.} \textbf{\bibinfo{volume}{A299}},
  \bibinfo{pages}{39} (\bibinfo{year}{1978}).

\bibitem[{\citenamefont{Karataglidis
  et~al.}(1995{\natexlab{b}})\citenamefont{Karataglidis, Dortmans, Amos, and
  de~Swiniarski}}]{Ka96}
\bibinfo{author}{\bibfnamefont{S.}~\bibnamefont{Karataglidis}},
  \bibinfo{author}{\bibfnamefont{P.~J.} \bibnamefont{Dortmans}},
  \bibinfo{author}{\bibfnamefont{K.}~\bibnamefont{Amos}}, \bibnamefont{and}
  \bibinfo{author}{\bibfnamefont{R.}~\bibnamefont{de~Swiniarski}},
  \bibinfo{journal}{Phys. Rev. C} \textbf{\bibinfo{volume}{53}},
  \bibinfo{pages}{838} (\bibinfo{year}{1995}{\natexlab{b}}).

\bibitem[{\citenamefont{Barker et~al.}(1981)\citenamefont{Barker, Smith,
  Morrison, and Amos}}]{Ba81}
\bibinfo{author}{\bibfnamefont{F.}~\bibnamefont{Barker}},
  \bibinfo{author}{\bibfnamefont{R.}~\bibnamefont{Smith}},
  \bibinfo{author}{\bibfnamefont{I.}~\bibnamefont{Morrison}}, \bibnamefont{and}
  \bibinfo{author}{\bibfnamefont{K.}~\bibnamefont{Amos}}, \bibinfo{journal}{J.
  Phys.} \textbf{\bibinfo{volume}{G7}}, \bibinfo{pages}{657}
  (\bibinfo{year}{1981}).

\bibitem[{\citenamefont{Petrovich and Love}(1983)}]{Pe83}
\bibinfo{author}{\bibfnamefont{F.}~\bibnamefont{Petrovich}} \bibnamefont{and}
  \bibinfo{author}{\bibfnamefont{W.~G.} \bibnamefont{Love}},
  \bibinfo{journal}{Nucl. Phys.} \textbf{\bibinfo{volume}{A354}},
  \bibinfo{pages}{499} (\bibinfo{year}{1983}).

\bibitem[{\citenamefont{Stricker et~al.}(1979)\citenamefont{Stricker, McManus,
  and Carr}}]{St79}
\bibinfo{author}{\bibfnamefont{K.}~\bibnamefont{Stricker}},
  \bibinfo{author}{\bibfnamefont{H.}~\bibnamefont{McManus}}, \bibnamefont{and}
  \bibinfo{author}{\bibfnamefont{J.~A.} \bibnamefont{Carr}},
  \bibinfo{journal}{Phys. Rev. C} \textbf{\bibinfo{volume}{19}},
  \bibinfo{pages}{929} (\bibinfo{year}{1979}).

\bibitem[{\citenamefont{Mihaila and Heisenberg}(2000)}]{Mi00}
\bibinfo{author}{\bibfnamefont{B.}~\bibnamefont{Mihaila}} \bibnamefont{and}
  \bibinfo{author}{\bibfnamefont{J.~H.} \bibnamefont{Heisenberg}},
  \bibinfo{journal}{Phys. Rev. C} \textbf{\bibinfo{volume}{61}},
  \bibinfo{pages}{54309} (\bibinfo{year}{2000}).

\bibitem[{\citenamefont{Sick and McCarthy}(1970)}]{Si70}
\bibinfo{author}{\bibfnamefont{I.}~\bibnamefont{Sick}} \bibnamefont{and}
  \bibinfo{author}{\bibfnamefont{J.~S.} \bibnamefont{McCarthy}},
  \bibinfo{journal}{Nucl. Phys.} \textbf{\bibinfo{volume}{A150}},
  \bibinfo{pages}{631} (\bibinfo{year}{1970}).

\bibitem[{\citenamefont{Raynal}(1998)}]{Ra98}
\bibinfo{author}{\bibfnamefont{J.}~\bibnamefont{Raynal}}
  (\bibinfo{year}{1998}), \bibinfo{note}{computer program DWBA98, NEA 1209/05}.

\bibitem[{\citenamefont{Carr et~al.}(1983)\citenamefont{Carr, Petrovich,
  Halderson, Holtkamp, and Cottingame}}]{Ca83}
\bibinfo{author}{\bibfnamefont{J.~A.} \bibnamefont{Carr}},
  \bibinfo{author}{\bibfnamefont{F.}~\bibnamefont{Petrovich}},
  \bibinfo{author}{\bibfnamefont{D.}~\bibnamefont{Halderson}},
  \bibinfo{author}{\bibfnamefont{D.~B.} \bibnamefont{Holtkamp}},
  \bibnamefont{and} \bibinfo{author}{\bibfnamefont{W.~B.}
  \bibnamefont{Cottingame}}, \bibinfo{journal}{Phys. Rev. C}
  \textbf{\bibinfo{volume}{27}}, \bibinfo{pages}{1636} (\bibinfo{year}{1983}).

\bibitem[{\citenamefont{Amos and Morrison}(1981)}]{Am81}
\bibinfo{author}{\bibfnamefont{K.}~\bibnamefont{Amos}} \bibnamefont{and}
  \bibinfo{author}{\bibfnamefont{I.}~\bibnamefont{Morrison}},
  \bibinfo{journal}{Phys. Rev. C} \textbf{\bibinfo{volume}{23}},
  \bibinfo{pages}{1679} (\bibinfo{year}{1981}).

\bibitem[{\citenamefont{Seifert et~al.}(1993)}]{Se93}
\bibinfo{author}{\bibfnamefont{H.}~\bibnamefont{Seifert}} \bibnamefont{et~al.},
  \bibinfo{journal}{Phys. Rev. C} \textbf{\bibinfo{volume}{47}},
  \bibinfo{pages}{1615} (\bibinfo{year}{1993}).

\bibitem[{\citenamefont{Amos et~al.}(1984)}]{Am84}
\bibinfo{author}{\bibfnamefont{K.}~\bibnamefont{Amos}} \bibnamefont{et~al.},
  \bibinfo{journal}{Nucl. Phys.} \textbf{\bibinfo{volume}{A413}},
  \bibinfo{pages}{255} (\bibinfo{year}{1984}).

\bibitem[{\citenamefont{Henderson et~al.}(1979)}]{He79}
\bibinfo{author}{\bibfnamefont{R.~S.} \bibnamefont{Henderson}}
  \bibnamefont{et~al.}, \bibinfo{journal}{Aust. J. Phys.}
  \textbf{\bibinfo{volume}{32}}, \bibinfo{pages}{411} (\bibinfo{year}{1979}).

\end{thebibliography}

\end{document}